\documentclass[12pt,a4paper]{article}

\usepackage{a4wide}
\usepackage{amsmath}
\usepackage{amssymb}
\usepackage{amsfonts}
\usepackage{epsfig}
\usepackage{exscale}
\usepackage{float}
\usepackage{bbm}
\usepackage[numbers,sort&compress]{natbib}

\makeatletter
\@addtoreset{equation}{section}
\makeatother

\title{
The  Solution of the Relativistic Schrodinger\\ Equation  for the $\delta'$-Function Potential \\in 1-dimension
Using Cutoff Regularization}

\author{M.\ H.\ Al-Hashimi$^{a,b}$, and  A.\ M.\ Shalaby$^{b,c}$
\footnote{Contact information: M.\ H.\ Al-Hashimi: hashimi@itp.unibe.ch,
+41 31 631 8878; A.\ Shalaby, amshalab@qu.edu.qa, +974 4403 4630.}
\\ \\
$^a$ Albert Einstein Center for Fundamental Physics \\
Institute for Theoretical Physics, Bern University \\
Sidlerstrasse 5, CH-3012 Bern, Switzerland \\ \\
$^b$ Department of Mathematics, Statistics, and Physics \\
Qatar University, Al Tarfa, Doha 2713, Qatar
\\ \\
$^c$ Physics Department, Faculty of Science \\
Mansoura University, Egypt \\ \\}

\begin{document}

\maketitle

\vspace{-1cm}

\begin{abstract} \normalsize
We study the relativistic version of Schr\"odinger equation for a point particle in 1-d with potential of the first derivative of the delta function. The momentum cutoff regularization is used to study the bound state and scattering states. The initial calculations show that the reciprocal of the bare coupling constant is ultra-violet divergent, and the resultant expression cannot be renormalized in the usual sense. Therefore a general procedure has been developed to derive different physical properties of the system.  The procedure is used first on the non-relativistic case for the purpose of clarification and comparisons. The results from the relativistic case show that this system behaves exactly like the delta function potential, which means it also shares the same features with quantum field theories, like being asymptotically free, and in the massless limit, it undergoes dimensional transmutation and it possesses an infrared conformal fixed point.
\end{abstract}
\section{Introduction}
Investigating a one particle relativistic quantum mechanical system proves to be a nontrivial process. Even for a free quantum mechanical relativistic point
we can get some interesting properties \cite{Bak73,Alm84,Str06,AlH09}. For example, it was shown that a minimal position-velocity wave
packet of a particle can spread in such a way that probability leaks out of the lightcone. Therefore, studying relativistic contact interactions is expected to give even more rich and interesting features.

In non-relativistic quantum mechanics, contact interactions have been studied
in great detail
\cite{Cal88,Tho79,Beg85,Hag90,Jac91,Fer91,Gos91,Mea91,Man93,Phi98}. Unlike the non-relativistic case, the
relativistic $\delta$-function potential gives rise to ultra-violet divergences, which is regularized
and renormalized using dimensional regularization. The approach is widely used in quantum field theories \cite{Bol72,Bol72a,tHo72,Bie14}. The need for any regularization method can be avoided by studying the problem as an application of the theory of the self-adjoint extensions of the pseudo-differential operators. This has already been investigated
in the mathematical literature, by using an abstract mathematical approach\cite{Alb97}. Recently, we studied the problem by directly solving the relativistic version of the Schr\"odinger equation of the Hamiltonian $H = \sqrt{p^2 + m^2}+\lambda\delta(x)$ in 1-d, where we used dimensional regularization to show that the system has remarkable features. For example, the relatively simple system shares many features with some complex quantum field theories, like asymptotic freedom, dimensional transmutation in the massless limit,  and it also possesses an infra-red conformal fixed point \cite{Our}. The same problem was studied using cutoff regularization \cite{Our2}. The solution gives the same results obtained using dimensional regularization.

The problem of the $\delta'$-function potential has attracted less attention than the $\delta$-function potential, that is aside from being studied generally in the context of contact interactions. It has been studied non-relativistically \cite{Hold,Albbook,Arnbak1,Zhao,Griff,Toy,Widmer,Fas013}, and in the context of Dirac equation \cite{Cal14,Arnbak11}. The main difference between the $\delta$-function potential and the $\delta'$-function potential is that the problem in the later needs to be regularized even in the 1-d non-relativistic case.

In this paper, we study  the relativistic version of the Schr\"odinger equation of the Hamiltonian $H = \sqrt{p^2 + m^2}+\lambda_1 \delta'(x)$ in 1-d using cutoff regularization. Normally, equivalent theories in quantum field theory are considered to be non-renormalizable, that is because the coupling constant has a positive power of length, and thus the theory is non-renormalizable by power counting \cite{Pes}. However, in this work it has been proven that the relativistic theory can be regularized.  The  non-relativistic $\delta'$-function potential problem has also been investigated in details for the sake of comparison, and to present a general method to treat this problem. Our results for  the $\delta'$-function potential shows the same remarkable properties of the $\delta$-function potential case. It  also shares several non-trivial features with relativistic quantum field theories. In particular, just like quantum chromodynamics (QCD) \cite{Fri73}, it is asymptotically free \cite{Gro73,Pol73}. On the other hand there is a subtle difference in the expression of the wave function from the $\delta$-function potential, which is a requirement for satisfying the boundary condition at the contact point.

In our previous work \cite{Our,Our2}, we proved that the bound state wave function takes the following form
\begin{eqnarray}\label{deltaWF}
\Psi_B(x) &=& \frac{\lambda \Psi_B(0)}{2 \pi} \int_{-\infty}^{\infty}dp \
\frac{\exp(i p x)}{E_B - \sqrt{p^2 + m^2}}\nonumber \\ &=&  \lambda \Psi_B(0) \left[\frac{1}{\pi}  \int_m^\infty d\mu
\frac{\sqrt{\mu^2 - m^2}}{E_B^2 - m^2 + \mu^2} \exp(- \mu |x|) +
\frac{E_B \exp(- \sqrt{m^2 - E_B^2} |x|)}{\sqrt{m^2 - E_B^2}}\right],\nonumber\\
\end{eqnarray}
where $E_B$ is the energy of the bound state. It is clear from eq.(\ref{deltaWF}) that the wave function is real up to a phase constant. The bound state wave function diverges logarithmically at the origin, nevertheless it is renormalizable. The normalization condition is
\begin{eqnarray}\label{NormCon}
&&\frac{2 \pi}{\lambda^2 |\Psi_B(0)|^2} = \frac{2 E_B}{m^2 - E_B^2} +
\frac{m^2}{(m^2 - E_B^2)^{3/2}}\left(\pi + 2 \arcsin\frac{E_B}{m}\right).
\end{eqnarray}
The above equation gives unusual statement; although $\Psi_B(0)$ is divergent, $\lambda\Psi_B(0)=C_1 $ is finite. This can be better understood in the context of the cutoff regularization \cite{Our2}. For this case, $\lambda$ can be obtained as a function of the cutoff momentum $\Lambda$;
\begin{equation}\label{lambS1}
    \lambda(\Lambda)=\frac{1}{I(E_B,\Lambda)},
\end{equation}
where
\begin{eqnarray}\label{lambS2}
I(E_B,\Lambda)=\frac{1}{2\pi }\left( \int_{-\Lambda }^{\Lambda }\frac{1}{E_B-%
\sqrt{p^{2}+m^{2}}}dp\right).
\end{eqnarray}
From eq(\ref{lambS2}) and eq(\ref{lambS1}), it is obvious that $\lambda(\Lambda)\rightarrow 0$ as $\Lambda \rightarrow \infty$. On the other hand  eq.(\ref{deltaWF}) gives
\begin{equation}\label{PsiSm}
  \Psi_B(0)=C_1 I(E_B),
\end{equation}
where
\begin{equation}\label{ILim}
    I(E_B)=\lim_{\Lambda\rightarrow \infty} I(E_B,\Lambda),
\end{equation}
and $C_1$ is a constant. Eq.(\ref{PsiSm}) means that $\Psi_B(0)\rightarrow \infty$ as $\Lambda \rightarrow \infty$. Under this framework, it is understandable how a vanishing quantity times a divergent quantity give a finite quantity that depends on the energy of the bound state and the mass. This concept, as basic as it is, is very important to understand the mathematical approach that we are using to study the present problem.
\section{The Non-Relativistic Case}
To understand  the relativistic $\delta ^{\prime }(x)$-function potential problem, it is important to study the non-relativistic solution using certain procedure of cutoff regularization. The Schr\"odinger equation in this case is
\begin{equation}\label{Schrodinger3}
\frac{p^{2}}{2m}\Psi (x)+\lambda_1 \delta ^{\prime }(x)\Psi (x)=\Delta E\Psi (x),
\end{equation}
where $\lambda_1$ is the bare coupling constant. In momentum space, the above equation is
\begin{equation}\label{pSpaceNon1}
\frac{p^{2}}{2m}\widetilde{\Psi }(p)+\lambda_1 \int_{-\infty }^{\infty
}\delta ^{\prime }(x)\Psi (x)e^{-ipx}dx=\Delta E\text{ }\widetilde{\Psi }(p),
\end{equation}
where
\begin{equation}
\int_{-\infty }^{\infty }\delta ^{\prime }(x)\Psi
(x)e^{-ipx}dx=-\int_{-\infty }^{\infty }\delta (x)\frac{d\Psi (x)}{dx}%
e^{-ipx}dx+ip\int_{-\infty }^{\infty }\delta (x)\Psi (x)e^{-ipx}dx.
\end{equation}
Therefore, eq.(\ref{pSpaceNon1}) can be written as
\begin{equation}\label{Schr1Deriv}
\frac{p^{2}}{2m}\widetilde{\Psi }(p)+\lambda_1 (ip\Psi (0)-\Psi ^{\prime
}(0))=\Delta E \widetilde{\Psi }(p),
\end{equation}
where
\begin{eqnarray}\label{FourierT}
\Psi(x) = \frac{1}{2 \pi} \int dp \ \widetilde \Psi(p) \exp(i p x), \quad
\Psi(0) = \frac{1}{2 \pi} \int dp \ \widetilde \Psi(p), \
\nonumber \\
\Psi^{\prime}(x) = \frac{1}{2 \pi} \int dp \ ip \widetilde \Psi(p) \exp(i p x), \quad
\Psi^{\prime}(0) = \frac{1}{2 \pi} \int dp \ ip \widetilde \Psi(p).
\end{eqnarray}
Accordingly, we can write eq.(\ref{pSpaceNon1}) as
\begin{equation}\label{Schr1DerivNon}
\frac{p^2}{2m}\widetilde{\Psi }(p)+\lambda_1 (ip\Psi (0)-\Psi ^{\prime
}(0))=\Delta E\widetilde{\Psi }(p).
\end{equation}
For the bound state, the wave function in coordinate space can be obtained using eq.(\ref{Schr1DerivNon})
\begin{equation}\label{PsideltaDerWFB}
\Psi_B (x)=\frac{m\lambda_1 }{\pi } \int_{-\infty }^{\infty }
\frac{ip\Psi (0)-\Psi ^{\prime}(0)}{2m\Delta E_B-p^2}e^{ipx}dp,
\end{equation}
where $\Delta E_B$ is the binding energy. The above equation can be written in a more compact form, that is
\begin{equation}\label{GeneralPsiNon2}
\Psi_B (x)=\lambda_1 \left(-I_{0}(x,\Delta E_B)\Psi_B ^{\prime}(0)+I_{1}(x,\Delta E_B)\Psi_B(0)\right),
\end{equation}
where
\begin{equation}\label{IkNon}
I_{k}(x,\Delta E_B)=\frac{m}{\pi }\left( P.V.\int_{-\infty }^{\infty }\frac{(ip)^{k}e^{ipx}%
}{2m\Delta E_B-p^2}dp\right) =\frac{\partial ^{k}I_{0}(x,\Delta E_B)}{\partial x^{k}
},
\end{equation}
From eq.(\ref{GeneralPsiNon2}), a general expression for wave function for the bound state in coordinate space can be obtained using contour integral (see Figure.1 top panel).
\begin{equation}\label{NonRelWFspace}
  \Psi_B (x)=\lambda_1\exp(-\sqrt{-2\Delta E_B m}|x|)\left(\Psi_B ^{\prime}(0) \sqrt{\frac{m}{-2 \Delta E_B}}+m \Psi_B (0)\text{sgn}(x)\right).
\end{equation}
The above expression can be considered as the unregularized expression of the wave function.
\begin{figure}[t]
\begin{center}
\epsfig{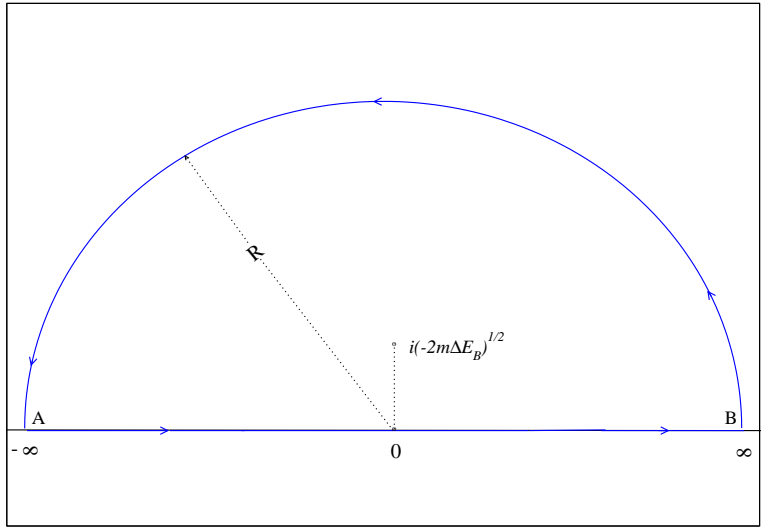} \vskip1.6 cm
\epsfig{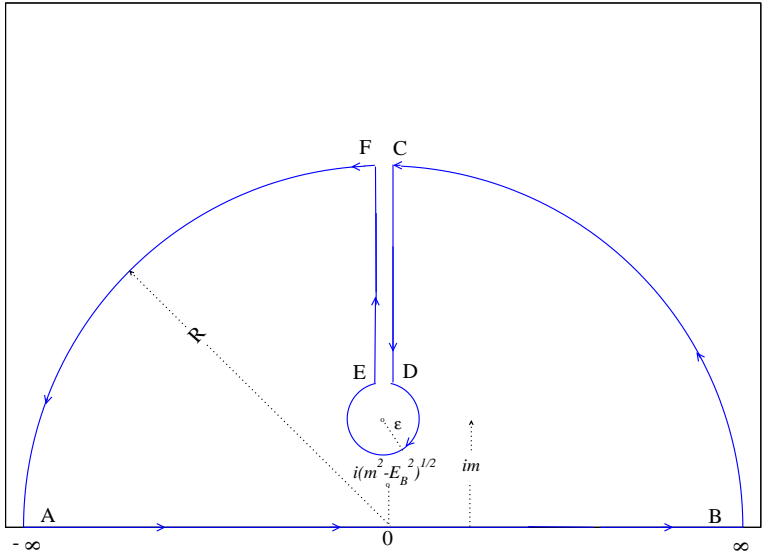}
\end{center}
\caption{\it The integration contours for obtaining the wave function of
the bound state. In the non-relativistic case, there is a pole inside the contour at $i\sqrt{-2m\Delta E_B}$, but no branch cut(top panel).
For relativistic case, there is a branch cut along the positive imaginary axis, starting at $p = im$, and there is a pole at $p = i \sqrt{m^2 - E_B^2}$ (bottom panel).}
\label{isofig2}
\end{figure}
From eq.(\ref{GeneralPsiNon2}) we get
\begin{equation}\label{gapfirstNon1}
    \Psi_B(0)=-\lambda_1 I_0(\Delta E_B) \Psi_B^{\prime}(0),
\end{equation}
\begin{equation}\label{gapfirstNon2}
    \Psi_B^{\prime}(0)=\lambda_1 I_2(\Delta E_B) \Psi_B(0),
\end{equation}
where $I_k(\Delta E_B)\equiv I_k(0,\Delta E_B)$. From the above two equations, we get the gap equation, that is
\begin{equation}\label{labdaNon1}
\frac{1}{\lambda_1}=\pm\sqrt{-I_0(\Delta E_B)I_2(\Delta E_B)}=\pm  \frac{1}{|\lambda_1|}.
\end{equation}
The right hand side of eq.(\ref{labdaNon1}) diverges. Therefore, the problem needs to be regularized. This can be done by regularizing the integrals $I_k(x,\Delta E_B)$. For cutoff regularization, the interval of the integral  in eq.(\ref{IkNon}) should be changed to $[-\Lambda,\Lambda]$, where $\Lambda$ is the cutoff momentum. We define $I_k(\Delta E_B,\Lambda)\equiv I_k(0,\Delta E_B,\Lambda)$, accordingly eq.(\ref{IkNon}) can be written as
\begin{equation}\label{IkNonCut}
I_{k}(\Delta E_B,\Lambda)=\frac{m}{\pi }\left( \int_{-\Lambda }^{\Lambda }\frac{(ip)^{k}%
}{2m\Delta E_B-p^2}dp\right).
\end{equation}
The values of the integrals $I_0(\Delta E_B,\Lambda)$ and $I_2(\Delta E_B,\Lambda)$ can be calculated from eq.(\ref{IkNonCut}), which gives
\begin{eqnarray}\label{INON1}
    I_0(\Delta E_B,\Lambda)&=&-\frac{1}{\pi}\sqrt{\frac{2m}{-\Delta E_B}}\arctan \left(\frac{\Lambda}{\sqrt{-2 \Delta E_B m}}\right),\nonumber\\
    I_2(\Delta E_B,\Lambda)&=&\frac{2m}{\pi}\left(\Lambda-\sqrt{-2m \Delta E_B}\arctan \left(\frac{\Lambda}{\sqrt{-2 \Delta E_B m}}\right)\right).
\end{eqnarray}
It is obvious from the above equations that $ I_0(\Delta E_B,\Lambda)$ is finite for $\Lambda \rightarrow\infty$. On the other hand $I_2(\Delta E_B,\Lambda)$ is linearly ultra-violet divergent. Therefore, we have
\begin{eqnarray}\label{INONc1}
    I_{0}(\Delta E_B)&=&-\lim_{\Lambda\rightarrow \infty}\frac{1}{\pi}\sqrt{\frac{2m}{-\Delta E_B}}\arctan \left(\frac{\Lambda}{\sqrt{-2 \Delta E_B m}}\right)=- \sqrt{\frac{m}{-2\Delta E_B}}.
\end{eqnarray}
In this context, the gap equation is
\begin{equation}\label{labdaNonCut1}
    \frac{1}{\lambda_1(\Lambda)}=\pm\sqrt{-I_0(\Delta E_B,\Lambda)I_2(\Delta E_B,\Lambda)},
\end{equation}
and the wave function in eq.(\ref{GeneralPsiNon2}) can be written as
\begin{equation}
    \Psi_B (x)=\lim_{\Lambda\rightarrow\infty}\frac{m\lambda_1 }{\pi } \int_{-\Lambda }^{\Lambda }
\frac{ip\Psi (0,\Lambda)-\Psi ^{\prime}(0,\Lambda)}{2m\Delta E_B-p^2}e^{ipx}dp,
\end{equation}
therefore we get
\begin{eqnarray}\label{GeneralPsiNon2Cut}
\Psi_B (x)&=&\lim_{\Lambda\rightarrow\infty}\Psi_B (x,\Lambda)\nonumber\\&=&\lim_{\Lambda\rightarrow\infty}\lambda_1 (\Lambda)\left[-I_{0}(x,\Delta E_B,\Lambda)\Psi_B ^{\prime}(0,\Lambda)+I_{1}(x,\Delta E_B,\Lambda)\Psi_B(0,\Lambda)\right],\nonumber \\
\end{eqnarray}
also we have
\begin{equation}\label{GeneralPsiDerivNon2Cut}
    \Psi^{\prime}_B (x)=\lim_{\Lambda\rightarrow\infty}\Psi^{\prime}_B (x,\Lambda).
\end{equation}
In addition,  eq.(\ref{GeneralPsiNon2Cut}) gives
 \begin{equation}\label{gapfirstNonCut1}
    \Psi_B(0,\Lambda)=-\lambda_1 I_0(\Delta E_B,\Lambda) \Psi_B^{\prime}(0,\Lambda),
\end{equation}
\begin{equation}\label{gapfirstNonCut2}
    \Psi_B^{\prime}(0,\Lambda)=\lambda_1 I_2(\Delta E_B,\Lambda) \Psi_B(0,\Lambda),
\end{equation}
therefore we get
\begin{equation}\label{PsiDiv}
    \Psi_B^{\prime}(0,\Lambda)=\pm \sqrt{-\frac{I_{2}(\Delta E_B,\Lambda)}{I_{0}(\Delta E_B,\Lambda)}}\Psi_B(0,\Lambda),
\end{equation}
where $\pm$ sign is correspond to $\lambda_1=\pm |\lambda_1|$. The above equation means that $\Psi_B^{\prime}(x)$ is singular at the origin as one can also verify this from eq.(\ref{PsideltaDerWFB}).

The normalization condition for the bound state imposes conditions on the values of $\lambda_1\Psi_B(0)$ and  $\lambda_1\Psi_B^{\prime}(0)$. From the  normalization condition, we have
\begin{equation}\label{NormCon}
    \int_{-\infty}^{\infty}|\Psi_B(x)|^2 dx=\lim_{\Lambda\rightarrow \infty}\frac{2m^2\lambda_1^2}{\pi} \int_{-\Lambda}^{\Lambda}\frac{p^2|\Psi_B(0)|^2+|\Psi_B^{\prime}(0)|^2}{(2m \Delta E_B-p^2)^2} dp=1.
\end{equation}
We can verify from eq.(\ref{PsideltaDerWFB}) that $\Psi_B(x)$ is real up to a phase constant, therefore the above equation can be written as
\begin{eqnarray}
 \int_{-\infty}^{\infty}|\Psi_B(x)|^2 dx&=&\lambda_1^2\Psi_B^{\prime}(0)^2 \frac{\sqrt{m}}{(-2\Delta E_B)^{3/2}}
 +\lambda_1^2\Psi_B(0)^2\frac{m^{3/2}}{\sqrt{-2 \Delta E_B}}\nonumber\\&=&\lambda_1^2\Psi_B^{\prime}(0)^2\left( \frac{\sqrt{m}}{(-2\Delta E_B)^{3/2}}-\frac{I_0(\Delta E_B)}{I_2(\Delta E_B)}\frac{m^{3/2}}{\sqrt{-2 \Delta E_B}}\right)\nonumber\\&=&1.
\end{eqnarray}
The above equation means that $\lambda_1\Psi_B^{\prime}(0) $ is a finite non-zero number, which is given by the following relation
\begin{equation}\label{PsiPValueNon}
   \lambda_1\Psi_B^{\prime}(0)=\pm\left( \frac{\sqrt{m}}{(-2\Delta E_B)^{3/2}}-\frac{I_0(\Delta E_B)}{I_2(\Delta E_B)}\frac{m^{3/2}}{\sqrt{-2 \Delta E_B}}\right)^{-1/2}.
\end{equation}
Since
\begin{equation}
  \lim_{\Lambda\rightarrow\infty}  \frac{I_0(\Delta E_B,\Lambda)}{I_2(\Delta E_B,\Lambda)}=\frac{I_0(\Delta E_B)}{I_2(\Delta E_B)}=0,
\end{equation}
we can write eq.(\ref{PsiPValueNon}) as
\begin{equation}\label{NonRelNor}
  \lambda_1\Psi_B^{\prime}(0)=\pm\frac{ (-2\Delta E_B )^{3/4}}{m^{1/4}}.
\end{equation}
The above equation together with eq.(\ref{gapfirstNonCut1}) and eq.(\ref{labdaNon1}) gives $\lambda_1 \Psi_B(0,\Lambda)\sim \Lambda^{-1/2}$, therefore we get
\begin{equation}\label{PsiLimit}
    \lim_{\Lambda\rightarrow \infty}\lambda_1(\Lambda) \Psi_B(0,\Lambda)=0.
\end{equation}
From previous discussion, the wave function for the bound state can finally be written  as
\begin{eqnarray}\label{NonRelWFfinal1}
    \Psi_B (x)&=&\pm\frac{\varkappa^{3/2}}{m}\Bigg(\frac{m}{\varkappa}\exp(-\varkappa|x|)\nonumber\\
    &\pm& \lim_{\Lambda\rightarrow \infty}\frac{m}{\pi}  \sqrt{-\frac{I_0(\Delta E_B)}{I_2(\Delta E_B,\Lambda)}}\int_{-\Lambda }^{\Lambda }\frac{ip e^{ipx}}{2m\Delta E_B-p^2}dp\Bigg),
\end{eqnarray}
where $\Delta E_B=-\varkappa^2/2m$. The $\pm$ sign inside the bracket in eq.(\ref{NonRelWFfinal1}) is correspond to $\lambda_1=\pm |\lambda_1|$. The second term in the bracket of  eq.(\ref{NonRelWFfinal1}) vanishes as $\Lambda\rightarrow\infty$, however it is very important to keep in mind that the term cannot be simply put to zero in the expression of  $\Psi^{\prime}_B (x)$. This can be better understood from eq.(\ref{GeneralPsiNon2}) that gives
\begin{equation}
     \Psi^{\prime}_B (x)=\lambda_1 \left(-I_{1}(x,\Delta E_B)\Psi_B ^{\prime}(0)+I_{2}(x,\Delta E_B)\Psi_B(0)\right),
\end{equation}
which means that we can no longer ignore the second term, because  $I_{2}(x,\Delta E_B)$ diverges at the origin such that $I_{2}(\Delta E_B)\Psi_B(0)\lambda_1\rightarrow \infty$. As a consequence of this argument, we can not say that the wave function in eq.(\ref{NonRelWFfinal1}) is even, but we can say that it has a diminishing odd part for any $x\in (-\infty,\infty)$. Only with such setting we can satisfy the boundary condition in eq.(\ref{gapfirstNon1}) and eq.(\ref{gapfirstNon2}) simultaneously.
\begin{figure}[t]
\begin{center}
\epsfig{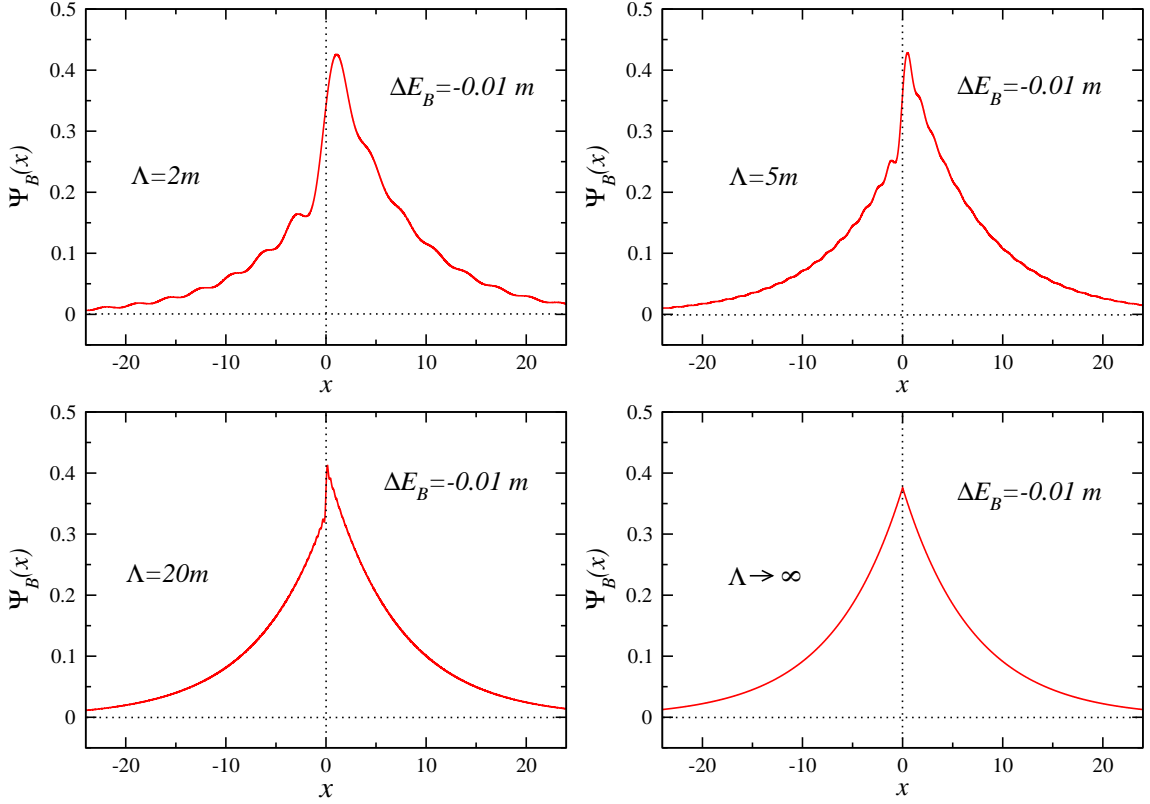}
\end{center}
\caption{\it Bound state wave function in coordinate space for the non-relativistic case. The value of the binding energy is $\Delta E_B =-0.01 m$,  with different values of $\Lambda=2m, 5m,20m$, and $\infty$. The wave function turns to an even function with the increase of $\Lambda$. }
\label{isofig3}
\end{figure}

Now we study the scattering wave function for the non-relativistic case. For this case we use the following ansatz
\begin{equation}\label{GeneralAnsatzNon}
  \tilde{\Psi}_E(p) =A\delta(p-\sqrt{2m\Delta E})+B\delta(p+\sqrt{2m\Delta E})+\tilde{\Phi}_E(p), \hspace{10mm} \Delta E=\frac{k^2}{2m},
\end{equation}
where $A$ and $B$ are arbitrary constants that will be defined later.
To calculate the scattering states, we must calculate $\Phi _{E}(x)$. Substituting for $\tilde{\Psi}_E(p)$  from eq.(\ref{GeneralAnsatzNon}) into eq.(\ref{Schr1DerivNon}), and then solving for   $\tilde{\Phi}_E(p)$ we get
\begin{equation}\label{PhiNonP}
\tilde{\Phi}_E(p)=\frac{2m \lambda_1
}{2m \Delta E-p^2}\left( ip\left( \frac{A+B}{2\pi }+\Phi _{E}(0)\right)
-\left( \left( \frac{A-B}{2\pi }\right) ik+\Phi _{E}^{\prime }(0)\right)
\right).
\end{equation}
In coordinate space
\begin{equation}\label{PhiNonX}
\Phi_{E}(x)=\frac{m\lambda_1}{\pi }\int_{-\infty }^{\infty }\frac{e^{ipx}dp%
}{2m \Delta E-p^2}\left( ip\left( \frac{A+B}{2\pi }+\Phi _{E}(0)\right)
-\left( \left( \frac{A-B}{2\pi }\right) ik+\Phi _{E}^{\prime }(0)\right)
\right),
\end{equation}
or in a more compact form
\begin{equation}\label{GeneralPhiNon2}
\Phi_E(x)=-\lambda_1I_{0}(x,\Delta E)\left(\Phi_E ^{\prime}(0)+ik\frac{A-B}{2\pi }\right)+\lambda_1 I_{1}(x,\Delta E)\left(\Phi_E(0)+\frac{A+B}{2\pi }\right),
\end{equation}
where the expression of $I_{k}(x,\Delta E)$ can be obtained from eq.(\ref{IkNon}) by replacing $\Delta E_B$ with $\Delta E$. Since $\Delta E>0$, we have
\begin{equation}
    I_{0}(\Delta E)=P.V.\int_{\infty}^{\infty}\frac{m}{\pi}\frac{1}{2m\Delta E-p^2}dp=0.
\end{equation}
\begin{figure}[t]
\begin{center}
\epsfig{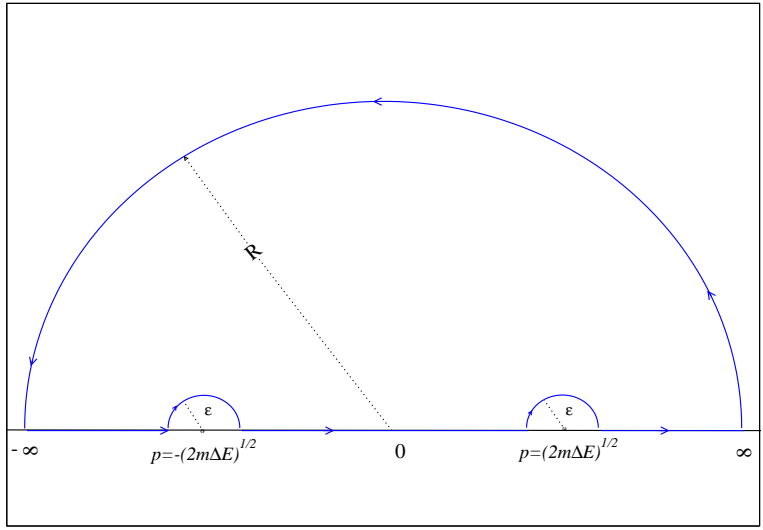} \vskip1.6 cm
\epsfig{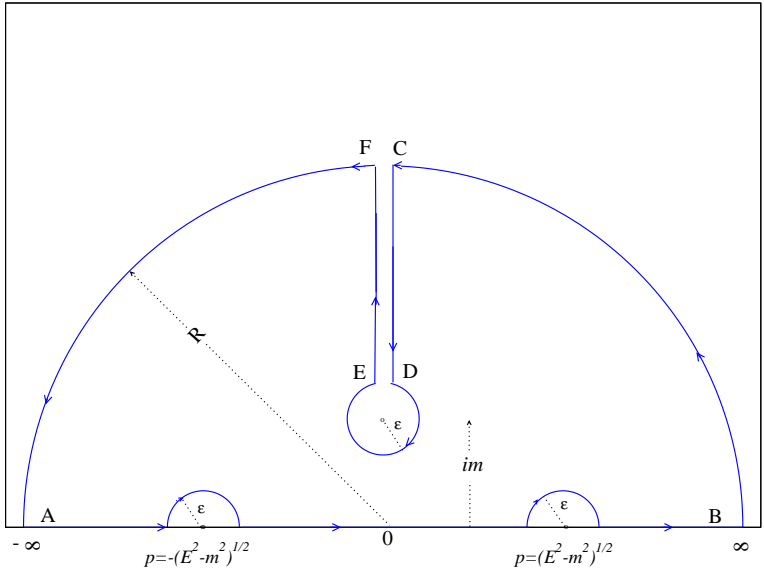}
\end{center}
\caption{\it The integration contours for obtaining the wave function of
scattering states. In the non-relativistic case, there are two  poles on the reals axis  at $\pm\sqrt{2m\Delta E}$, but no branch cut(top panel).
For relativistic case, there is a branch cut along the positive imaginary axis, starting at $p = im$, and there are two  poles on the real axis at $p = \pm\sqrt{E^2 - m^2}$ (bottom panel).}
\label{isofig2}
\end{figure}
From eq.(\ref{GeneralPhiNon2}) we get
\begin{equation}\label{ScattfirstNon1}
    \Phi_E(0)=-\lambda_1I_{0}(\Delta E)\left(\Phi_E ^{\prime}(0)+ik\frac{A-B}{2\pi }\right),
\end{equation}
\begin{equation}\label{ScattfirstNon2}
    \Phi_E^{\prime}(0)=\lambda_1 I_{2}(\Delta E)\left(\Phi_E(0)+\frac{A+B}{2\pi }\right).
\end{equation}
From eq.(\ref{GeneralPhiNon2}), and using momentum cutoff  we get
\begin{equation}\label{ScatfirstNon1}
    \Phi_E(0,\Lambda)=-\lambda_1I_{0}(\Delta E,\Lambda))\left(\Phi_E ^{\prime}(0,\Lambda)+ik\frac{A-B}{2\pi }\right),
\end{equation}
\begin{equation}\label{ScatfirstNon2}
    \Phi_E^{\prime}(0,\Lambda)=\lambda_1 I_{2}(\Delta E,\Lambda)\left(\Phi_E(0,\Lambda)+\frac{A+B}{2\pi }\right).
\end{equation}
Again here, we have $\Phi_E(x)=\lim_{\Lambda\rightarrow\infty}\Phi_E(x,\Lambda)$, and $\Phi_E^{\prime}(x)=\lim_{\Lambda\rightarrow\infty}\Phi_E^{\prime}(x,\Lambda)$. For $\Lambda\rightarrow\infty$, the above two equations give
 \begin{eqnarray}\label{Phi0Non2}
   -\lambda_1 \left(\frac{A-B}{2\pi}ik+\Phi^{\prime} _{E}(0)\right)&=&\frac{A+B}{2\pi(I_{0}(\Delta E_B)-I_{0}(\Delta E))}, \nonumber\\ \lambda_1(\Lambda) \left(\frac{A+B}{2\pi}+\Phi _{E}(0,\Lambda)\right)&\sim & \sqrt{\frac{1}{\Lambda}}.
\end{eqnarray}
Therefore we get
\begin{eqnarray}\label{PhideltaDerxFinal}
 \Phi _{E}(x)&=& (A+B)\Bigg(\frac{m}{2\pi^2I_0(\Delta E_B)}  \int_{-\infty }^{\infty }\frac{e^{ipx}dp}{2m \Delta E-p^2}\nonumber\\&\pm&\lim_{\Lambda\rightarrow\infty}\sqrt{\frac{1}{\Lambda}}\sqrt{\frac{m}{-8\pi^3I_0(\Delta E_B)}}\int_{-\Lambda }^{\Lambda}\frac{i p \ e^{ipx}dp}{2m \Delta E-p^2}\Bigg), \hspace{6mm} \lambda_1=\pm |\lambda_1|.
\end{eqnarray}
From the above two equations and eq.(\ref{GeneralPhiNon2}) we get
\begin{eqnarray}\label{PsiXdeltaFirstNon}
\Psi_E(x)&=& A e^{ikx} +B e^{-ikx}+
\frac{(A+B)}{ I_0(\Delta E_B)} \frac{m \sin(k |x|)}{k} \nonumber \\ &\pm &\lim_{\Lambda\rightarrow\infty}(A+B)\sqrt{\frac{m}{-2\pi I_0(\Delta E_B)}} \sqrt{\frac{1}{\Lambda}}\int_{-\Lambda }^{\Lambda }\frac{i p \ e^{ipx} dp}{k^2-p^2}, \nonumber\\  \lambda_1 &=& \pm |\lambda_1|.
\end{eqnarray}
We define the  coupling constant as
\begin{equation}\label{LambdaEB}
    \lambda(\Delta E_B)\equiv \lambda= \frac{1}{I_0(\Delta E_B)}= \sqrt{\frac{-2\Delta E_B}{m}}\Rightarrow  \ \ \ \Delta E_B=-\frac{m \lambda^2 }{2},
\end{equation}
and therefore, when we remove the cutoff, the scattering wave function can be written as 
\begin{eqnarray}\label{PsiXdeltaNon}
\Psi_E(x)&=& A e^{ikx} +B e^{-ikx}+
\lambda (A+B) \frac{m}{k} \sin(k |x|),\ \ \ \ \Delta E= \frac{k^2}{2m}.
\end{eqnarray}
It is important to mention here that for both case $\lambda_1 = |\lambda_1|$ and $\lambda_1 = -|\lambda_1|$, we get attractive scattering states given by eq.(\ref{PsiXdeltaNon}) with $ \lambda(\Delta E_B)<0$. This means that the regularization does not lead to a repulsive $\delta'$-function potential. Like in the case of the bound state,  we can not say that $\Phi_E(x)$ is an even function, but we can say that it has a diminishing odd part for any value of $x\in (-\infty,\infty)$.

The reflected wave function in the region $I$ to the left of the contact point, and transmitted wave function in the region $II$ to the right of the contact point are defined as
\begin{equation}\label{ReflDelNon}
\Psi_I(x) = \exp(i k x) + R(k) \exp(- i k x),\ \ \ \Psi_{II}(x) = T(k) \exp(i k x).
\end{equation}
From the above two equations and from eq.(\ref{PsiXdeltaNon}) we get
\begin{eqnarray}
R(k) = - \frac{im\lambda}
{k + i m \lambda}, \quad
T(k) = \frac{k}{k + i m \lambda},\nonumber\\
A= \frac{2k + i m \lambda}{2k + 2i m \lambda}, \ \ \ \ \quad B=- \frac{i m \lambda}{2k + 2i m \lambda}\Rightarrow A+B=T(k).
\end{eqnarray}
Accordingly, eq.(\ref{PsiXdeltaNon}) can be written as
\begin{eqnarray}\label{PsiXdeltaNon2}
\Psi_E(x)&=& A e^{ikx} +B e^{-ikx}+
 T(k) \frac{m}{k} \sin(k |x|),\ \ \ \,
\end{eqnarray}

To prove that the resulting system is self-adjoint, we have to prove that the scalar product of the bound state with a scattering state vanishes, or
\begin{equation}\label{ScNonEEB}
    \langle \Psi_B|\Psi_E\rangle=0,\ \ \ \ \ \ \ \lambda_1= \pm|\lambda_1|,
\end{equation}
also we must prove that the scalar product of a scattering state with energy $E'$ with another scattering state with energy $E$ gives
\begin{equation}\label{ScNonEpE}
\langle \Psi_{E'}|\Psi_E\rangle \sim
\delta(\sqrt{2m\Delta E} - \sqrt{2m\Delta E'}),\ \ \ \ \ \ \ \ \  \  \lambda_1=\pm |\lambda_1|.
\end{equation}
This has been proved in details in appendix B.
\section{The Bound State of the Relativistic Problem }
The relativistic time-independent Schr\"odinger equation for the $\delta^{\prime}$-function potential is
\begin{equation}\label{Schrodinger3}
\sqrt{p^{2}+m^{2}}\Psi (x)+\lambda_1 \delta ^{\prime }(x)\Psi (x)=E\psi (x).
\end{equation}
In momentum space, the above equation takes the following form
\begin{equation}\label{Schr1Deriv}
\sqrt{p^{2}+m^{2}}\widetilde{\Psi }(p)+\lambda_1 (ip\Psi (0)-\Psi ^{\prime
}(0))=E\widetilde{\Psi }(p).
\end{equation}
For the bound state, the above equation gives
\begin{equation}\label{PsideltaDerWF}
\Psi_B (x)=\frac{\lambda_1 }{2\pi }\int_{-\infty }^{\infty }\frac{ip\Psi (0)-\Psi^{\prime}(0)}{E_B-\sqrt{p^{2}+m^{2}}}e^{ipx}dp.
\end{equation}
It is obvious from eq.(\ref{PsideltaDerWF}) that $\Psi_B(x)$ is real up to a phase constant. The above equation can be written in a more compact form, that is
\begin{equation}\label{PsideltaDerWFCom}
\Psi_B(x)=\lambda_1 (-\Psi ^{\prime }_B(0)I_0(x,E_B)+\Psi_B(0)I_1(x,E_B)),
\end{equation}
where $I_k(x,E_B)$ is defined as
\begin{equation}
I_{k}(x,E_B)=\frac{1}{2\pi }\left( \int_{-\infty }^{\infty }\frac{(ip)^{k}e^{ipx}%
}{E_B-\sqrt{p^{2}+m^{2}}}dp\right) =\frac{\partial ^{k}I_{0}(x,E_B)}{\partial x^{k}
}.
\end{equation}
From the expression of $\Psi(x)$ in eq.(\ref{PsideltaDerWF}), and the definition of $I_k(x,E_B)$, it is possible to express the wave function for the bound state in terms of the K-Bessel functions. This can be done by calculating first $I_0(x,E_B)$ in terms of the K-Bessel functions
\begin{eqnarray}\label{DeltaBessel}
 I_0(x,E_B)=\frac{1}{2\pi }\int_{-\infty }^{\infty
}\sum_{n=1}^{\infty }\frac{-E_B^{n-1}}{\left( p^{2}+m^{2}\right) ^{n/2}}%
e^{ipx}dp\nonumber\\=-\frac{1}{\sqrt{\pi}} \sum_{n=0}^\infty \left(\frac{E_B}{m}\right)^n
\left(\frac{m |x|}{2}\right)^{n/2}
\frac{K_{n/2}(m |x|)}{\Gamma\left(\frac{n+1}{2}\right)}.
\end{eqnarray}
From eq.(\ref{DeltaBessel}) and eq.(\ref{PsideltaDerWF}), we get
\begin{eqnarray}\label{PsideltaDerWFBes}
\Psi_B (x)&=&\frac{\lambda_1}{\sqrt{\pi}} \sum_{n=0}^\infty \Bigg(\left(\frac{E_B}{m}\right)^n
\left(\frac{m |x|}{2}\right)^{n/2}
\nonumber\\ &\times&\frac{\left(K_{n/2}(m |x|)\Psi^{\prime }_B(0)+\text{sgn}(x)K_{n/2-1}(m |x|)\Psi_B(0)\right)}{\Gamma\left(\frac{n+1}{2}\right)}\Bigg).
\end{eqnarray}
For a bound state $0 < E_B < m$, or strongly bound state $-m < E_B < 0$, the above series converges. On the other hand, in the case of the ultra-strong bound state when $E_B<-m$, the series diverges.

It is possible to obtain the wave function for all cases, bound, strongly bound, and ultra-strong bound, using the elegant contour integral method \cite{Our}. Again, here it is sufficient to calculate  $I_{0}(x,E_B)$ in order to calculate the wave function.  For the bound state $0 < E_B < m$, the contour has one pole inside the upper half circle at $p = i \sqrt{m^2 - E_B^2}$, and it also has a branch
cut along the positive imaginary axis starting at $p = i m$ as it is illustrated in Figure.1 top panel. In this case $I_{0}(x,E_B)$ is
\begin{equation}\label{I-boundstate}
I_{0}(x,E_B)= - \frac{1}{\pi} \int_m^\infty d\mu
\frac{\sqrt{\mu^2 - m^2}}{E_B^2 - m^2 - \mu^2} \exp(- \mu |x|) -
\frac{E_B \exp(- \sqrt{m^2 - E_B^2} |x|)}{\sqrt{m^2 - E_B^2}}.
\end{equation}
Accordingly, the wave function is
\begin{eqnarray}\label{boundstateDelta1}
\Psi_B(x) = \lambda_1 \Psi^{\prime}_B(0)\left(\frac{1}{\pi}\int_m^\infty d\mu
\frac{\sqrt{\mu^2 - m^2}}{E_B^2 - m^2 + \mu^2} \exp(- \mu |x|) +
\frac{E_B \exp(- \sqrt{m^2 - E_B^2} |x|)}{\sqrt{m^2 - E_B^2}}\right)\nonumber\\
-\lambda_1\Psi_B(0)\text{sgn}(x)\left(\frac{1}{\pi}\int_m^\infty d\mu
\frac{\mu\sqrt{\mu^2 - m^2}}{E_B^2 - m^2 + \mu^2} \exp(- \mu |x|) +
E_B \exp(- \sqrt{m^2 - E_B^2} |x|)\right).
\end{eqnarray}
For ultra-strong bound states, the pole is outside the contour, therefore we get
\begin{equation}\label{I-boundstateUltra}
I_{0}(x,E_B)= - \frac{1}{\pi} \int_m^\infty d\mu
\frac{\sqrt{\mu^2 - m^2}}{E_B^2 - m^2 - \mu^2} \exp(- \mu |x|),
\end{equation}
as a result, the wave function for this case is
\begin{eqnarray}\label{UltraBoundDeltaDriv}
\Psi_B(x)& =& \lambda_1\Psi^{\prime}_B(0)\left(\frac{1}{\pi}\int_m^\infty d\mu
\frac{\sqrt{\mu^2 - m^2}}{E_B^2 - m^2 + \mu^2} \exp(- \mu |x|)\right)\nonumber\\
&-&\lambda_1\Psi_B(0)\text{sgn}(x)\left(\frac{1}{\pi}\int_m^\infty d\mu
\frac{\mu\sqrt{\mu^2 - m^2}}{E_B^2 - m^2 + \mu^2} \exp(- \mu |x|)\right).
\end{eqnarray}
It is important to note here that the strongly bound and ultra-strong bound states have no equivalence in the non-relativistic solution. They are pure relativistic states. Their expression in  eq.(\ref{UltraBoundDeltaDriv}) is a result of the contribution of the branch cut in Figure.1. As we already know, the branch cut does not exist in the contour of the non-relativistic case.
Again here, the expressions for $\Psi_B(x)$ in eq.(\ref{boundstateDelta1}) and eq.(\ref{UltraBoundDeltaDriv}) are unregularized expressions of the bound state.

Like the non-relativistic case, the wave function in eq.(\ref{UltraBoundDeltaDriv}) is not normalizable because of the second term. In general, the expression of the wave function for $\delta^{\prime}$-function potential is  not normalizable. That is because the integral
\begin{equation}\label{DivNor}
    \int_{-\infty}^{\infty}\frac{p^{2}}{(E_B-\sqrt{p^2+m^2})^2}\rightarrow \infty.
\end{equation}
However, using the momentum cutoff regularization changes this, and the wave function can indeed be normalized, as we will explain later.
\section{The Gap Equation and Cutoff Regularization for the Relativistic Case }
First, we define
\begin{eqnarray}\label{Iinfnit}
I_k(E_B)=I_k(0,E_B)=\frac{1}{2\pi }P.V.\left( \int_{-\infty }^{\infty }\frac{(ip)^{k}}{E_B-%
\sqrt{p^{2}+m^{2}}}dp\right).
\end{eqnarray}
For odd $k$, and using eq.(\ref{Iinfnit}) we get
\begin{equation}
   I_{2j+1}(E_B)=0,\hspace{12mm} j=0,1,2,3,...
\end{equation}
The gap equation can be derived from the expression of $\Psi_{B}(x)$ and $\Psi_{B}^{\prime}(x)$ in  eq.(\ref{PsideltaDerWFCom}) at $x=0$, which gives
\begin{equation}\label{gapfirst1}
    \Psi_B(0)=-\lambda_1 I_0(E_B) \Psi_B^{\prime}(0),
\end{equation}
 \begin{equation}\label{gapfirst2}
    \Psi_B^{\prime}(0)=\lambda_1 I_2(E_B) \Psi_B(0).
\end{equation}
By solving  eq.(\ref{gapfirst1}) and eq.(\ref{gapfirst2}) for $\lambda_1$, we get
\begin{equation}\label{lambda1}
    \frac{1}{\lambda_1}=\pm\sqrt{-I_0(E_B)I_2(E_B)}.
\end{equation}
The right hand side of eq.(\ref{lambda1}) diverges, and therefore it must be regularized. This can be done by regularizing the integrals $I_k(E_B)$. For cutoff regularization, the interval of the integral  in eq.(\ref{Iinfnit}) should be changed to $[-\Lambda,\Lambda]$. Accordingly, the gap equation can be written as
\begin{equation}\label{lambdaCut1}
    \frac{1}{\lambda_1(\Lambda)}=\pm\sqrt{-I_0(E_B,\Lambda)I_2(E_B,\Lambda)},
\end{equation}
where
\begin{eqnarray}\label{IinfnitCut}
I_k(E_B,\Lambda)=\frac{1}{2\pi }P.V.\left( \int_{-\Lambda }^{\Lambda}\frac{(ip)^{k}}{E_B-%
\sqrt{p^{2}+m^{2}}}dp\right).
\end{eqnarray}
In addition, eq.(\ref{gapfirst1}) and eq.(\ref{gapfirst2}) can be written as
\begin{equation}\label{gapfirstCut1}
    \Psi_B(0,\Lambda)=-\lambda_1 I_0(E_B,\Lambda) \Psi_B^{\prime}(0,\Lambda),
\end{equation}
 \begin{equation}\label{gapfirstCut2}
    \Psi_B^{\prime}(0,\Lambda)=\lambda_1 I_2(E_B,\Lambda) \Psi_B(0,\Lambda).
\end{equation}
To evaluate $I_0(E_B,\Lambda)$, the right hand side of eq.(\ref{IinfnitCut}) for $k=0$ is expanded in powers of $E_B/\sqrt{p^2 + m^2}$. This gives
\begin{equation}
I_0(E_B,\Lambda) = - \frac{1}{2\pi} \int_{-\Lambda} ^{\Lambda} \left(
\frac{1}{\sqrt{p^2 + m^2}}+
\sum_{n=2}^\infty \left(\frac{E_B}{\sqrt{p^2 + m^2}}\right)^n\right)dp.
\end{equation}
If we take the limit $\Lambda\rightarrow \infty$, we find that all the terms in the summation are finite.  On the other hand, the first term is
logarithmically ultra-violet divergent. All the rest of the terms can be integrated separately when $\Lambda\rightarrow \infty$
and then re-summed. The summation is convergent for a bound state $0<E_B<m$ and a strong bound states $0>E_B>-m$ as it was explained in \cite{Our}, and we get
\begin{equation}
\label{IEB}
I_0(E_B,\Lambda) = \frac{1}{2\pi }
\log\left(\frac{\sqrt{\Lambda^2+m^2}-\Lambda}{\sqrt{\Lambda^2+m^2}+\Lambda}\right)- \frac{E_B}{2 \pi \sqrt{m^2 - E_B^2}}
\left(\pi + 2 \arcsin\frac{E_B}{m}\right).
\end{equation}
We denote the finite part of $I_k(E_B)$ as $I_{kc}(E_B)$. For example
\begin{equation}
\label{IEBc0}
I_{0c}(E_B) = -
 \frac{E_B}{2 \pi \sqrt{m^2 - E_B^2}}
\left(\pi + 2 \arcsin\frac{E_B}{m}\right),
\end{equation}
For an ultra-strong bound state with energy $E_B < - m$, the series  diverges. Still, the result
can be obtained by directly integrating the convergent expression, and taking the limit $\Lambda\rightarrow \infty$
\begin{eqnarray}
\label{IEBU}
I_{0c}(E_B)&=&\frac{1}{2 \pi} \int dp \ \left(\frac{1}{E_B - \sqrt{p^2 + m^2}} +
\frac{1}{\sqrt{p^2 + m^2}} \right) = \nonumber \\
&&\frac{E_B}{\pi \sqrt{E_B^2 - m^2}} \
\text{arctanh} \left(\frac{\sqrt{E_B^2 - m^2}}{E_B}\right), \hspace{6mm} E_B<-m.
\end{eqnarray}
The expression of $I_{2}(E_B,\Lambda)$ can be obtained similarly, and we get
\begin{eqnarray}
\label{IEB2}
I_2(E_B,\Lambda) &=& \frac{1}{2\pi }\Bigg(\Lambda\sqrt{\Lambda^2+m^2}+
\frac{m^2-2E_B^2}{2}\log\left(\frac{\sqrt{\Lambda^2+m^2}-\Lambda}{\sqrt{\Lambda^2+m^2}+\Lambda}\right)+2E_B\Lambda\Bigg)\nonumber\\&+&I_{2c}(E_B),
\end{eqnarray}
where $I_{2c}(E_B)$ is the finite part of $I_2(E_B)$. It is needless to calculate the value of $I_{2c}(E_B)$ because it will never play a role in the calculations, as we will show this later. It is obvious that $I_{2}(E_B)$ is $\Lambda^2$ ultra-violet divergent.

It is important to compare the order of smallness of $\lambda_1\Psi_B(0)$ relative to  $\lambda_1\Psi_B^{\prime}(0)$. From  eq.(\ref{gapfirstCut1}) and  eq.(\ref{gapfirstCut2}) we get
\begin{equation}\label{PsiDiv}
    \Psi_B^{\prime}(0,\Lambda)=\pm\sqrt{-\frac{I_{2}(E_B,\Lambda)}{I_{0}(E_B,\Lambda)}}\Psi_B(0,\Lambda),\ \ \ \ \ \lambda_1=\pm |\lambda_1|.
\end{equation}
It is true that both of $\Psi^{\prime}_B(x)$ and $\Psi_B(x)$ are singular at the origin, as one can verify from eq.(\ref{UltraBoundDeltaDriv}). However, eq.(\ref{PsiDiv}) means that $\Psi_B^{\prime}(0)$ is even more divergent than $\Psi_B(0)$. The normalization condition for the bound state can give the values of $\lambda_1\Psi_B(0)$ and  $\lambda_1\Psi^{\prime}_B(0)$. From the  normalization condition, we have
\begin{equation}\label{NormCon}
    \int_{-\infty}^{\infty}|\Psi_B(x)|^2 dx=\lim_{\Lambda\rightarrow \infty}\lambda_1^2 \int_{-\Lambda}^{\Lambda}\frac{p^2\Psi_B(0,\Lambda)^2+\Psi^{\prime}_B(0,\Lambda)^2}{(E_B-\sqrt{p^2+m^2})^2} dp=1,
\end{equation}
by using eq.(\ref{PsiDiv}), the above equation can be written as
\begin{eqnarray}
 \int_{-\infty}^{\infty}|\Psi_B(x)|^2 dx&=& \lim_{\Lambda\rightarrow \infty}\frac{1}{2\pi}\lambda_1^2\Psi^{\prime}_B(0,\Lambda)^2\Bigg(\frac{2 E_B}{m^2 - E_B^2} +
\frac{m^2}{(m^2 - E_B^2)^{3/2}}\left(\pi + 2 \arcsin\frac{E_B}{m}\right)\nonumber \\&-&\frac{I_{0}(E_B,\Lambda)}{I_{2}(E_B,\Lambda)}G(\Lambda)\Bigg)=1,
\end{eqnarray}
where
\begin{equation}\label{REq}
  G(\Lambda)=\int_{-\Lambda}^{\Lambda}\frac{p^2}{(E_B-\sqrt{p^2+m^2})^2} dp.
\end{equation}
By obtaining the above integral, we find that the expression of $G(\Lambda)$ diverges like $\Lambda$. In order that the bound state to be normalizable, we must have
\begin{eqnarray}\label{Psi0PPValue}
   \lambda_1\Psi_B^{\prime}(0)&=&\pm \sqrt{2\pi}\lim_{\Lambda\rightarrow \infty}\Bigg(\frac{2 E_B}{m^2 - E_B^2} +
\frac{m^2}{(m^2 - E_B^2)^{3/2}}\left(\pi + 2 \arcsin\frac{E_B}{m}\right)\nonumber \\&-&\frac{I_{0}(E_B,\Lambda)}{I_{2}(E_B,\Lambda)}G(\Lambda)\Bigg)^{-1/2}.
\end{eqnarray}
For $\Lambda\rightarrow \infty$, the term $(I_{0}(E_B,\Lambda)/I_{2}(E_B,\Lambda)) \ G(\Lambda)$  vanishes in the above equation, and we get
\begin{eqnarray}\label{Psi0PPValueF}
  C_1= \lambda_1\Psi_B'(0)&=&\pm \sqrt{2\pi}\Bigg(\frac{2 E_B}{m^2 - E_B^2} +
\frac{m^2}{(m^2 - E_B^2)^{3/2}}\left(\pi + 2 \arcsin\frac{E_B}{m}\right)\Bigg)^{-1/2},
\end{eqnarray}
this means that for this case, the wave function in eq.(\ref{PsideltaDerWF}) is normalizable, and $ \lambda_1\Psi_B'(0)$ is a finite number. According, the wave function in eq.(\ref{PsideltaDerWFCom}) can be written as
\begin{equation}\label{BRState}
 \Psi_B(x)=C_1\lim_{\Lambda\rightarrow \infty}\left(-I_{0}(x,E_B,\Lambda)\pm\sqrt{-\frac{I_{0}(E_B,\Lambda)}{I_{2}(E_B,\Lambda)}}I_{1}(x,E_B,\Lambda)\right),
\end{equation}
where $I_{0}(E_B,\Lambda)$ is given by eq.(\ref{I-boundstate}) for bound and strong bound states, while it is given by eq.(\ref{I-boundstateUltra}) for the ultra-bound state. As for the second term in the above equation, it can be proved that $I_{1}(x,E_B,\Lambda)$ has a pulse with peaks at $x=\pm a(E_B) \Lambda^{-1}$ near the origin, where $a(E_B)$ is a constant, also we get
\\$I_{1}(\pm a/\Lambda,E_B,\Lambda)\pm \sim b(E_B) \Lambda$, where $b(E_B)$ is another constant. Nevertheless, the pulse is suppressed by the term $\sqrt{I_{0}(E_B,\Lambda)/I_{2}(E_B,\Lambda)}$ as  $\Lambda\rightarrow \infty$. Therefore, after taking the limit $\Lambda\rightarrow \infty$, the second term can be neglected in comparison with the first term for any value of $x\in (-\infty,\infty)$.
This issue has been discussed in more details in appendix A. Like the non-relativistic case, we can not simply say that the second term is zero, because we can not neglect this term in the expression of $\Psi_B^{\prime}(x)$.
\begin{figure}[t]
\begin{center}
\epsfig{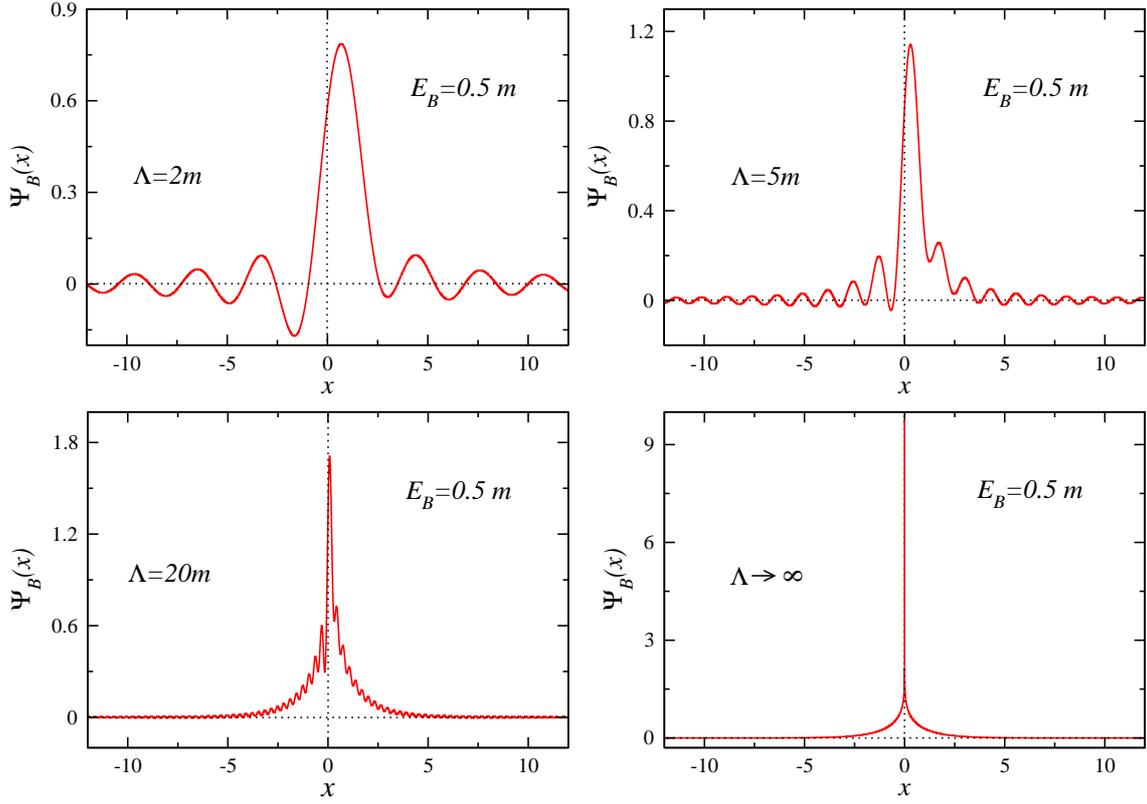}
\end{center}
\caption{\it Bound state wave function in coordinate space for the relativistic case with
$E_B = m/2$, and different values of $\Lambda=2m, 5m,20m$, and $\infty$. The wave function turns to an even function when $\Lambda\rightarrow\infty$.}
\label{isofig3}
\end{figure}
\section{The Scattering States for the Relativistic Case}
For the scattering states, we have $E>m$. A suitable ansatz for this case must be neither even or odd. Therefore, we assume that the solution has the following form
\begin{equation}\label{GeneralAnsatz}
  \tilde{\Psi}_E(p) =A\delta(p-\sqrt{E^{2}-m^{2}})+B\delta(p+\sqrt{E^{2}-m^{2}})+\tilde{\Phi}_E(p),
\end{equation}
By substituting eq.(\ref{GeneralAnsatz}) into eq.(\ref{Schr1Deriv}), we get
\begin{eqnarray}\label{PhiPdeltaD}
  ( \sqrt{p^2+m^2}-E)\tilde{\Phi}_E(p)+\left( ip\left( \frac{A+B}{2\pi }+\Phi _{E}(0)\right)
-\left( \left( \frac{A-B}{2\pi }\right) ik+\Phi _{E}^{\prime }(0)\right)
\right)=0\nonumber\\ \Rightarrow\tilde{\Phi}_E(p)=\frac{1
}{E-\sqrt{p^{2}+m^{2}}}\left( ip\left( \frac{A+B}{2\pi }+\Phi _{E}(0)\right)
-\left( \left( \frac{A-B}{2\pi }\right) ik+\Phi _{E}^{\prime }(0)\right)
\right).&&\nonumber\\
\end{eqnarray}
From eq.(\ref{PhiPdeltaD}), we get $\tilde{\Phi}_E(p)$ in coordinate space, that is
\begin{equation}
\Phi _{E}(x)=\frac{\lambda_1}{2\pi }\int_{-\infty }^{\infty }\frac{e^{ipx}dp%
}{E-\sqrt{p^{2}+m^{2}}}\left( ip\left( \frac{A+B}{2\pi }+\Phi _{E}(0)\right)
-\left( \left( \frac{A-B}{2\pi }\right) ik+\Phi _{E}^{\prime }(0)\right)
\right).
\end{equation}
We can write the above equation in a more compact form, that is
\begin{equation}\label{PhideltaDerWFCom}
\Phi_E(x)=\lambda_1\left(-\left(\frac{(A-B)ik}{2\pi }+\Phi _{E}^{\prime }(0)\right)I_0(x,E)+ \left( \frac{A+B}{2\pi }+\Phi _{E}(0)\right)I_1(x,E)\right).
\end{equation}
From eq.(\ref{PhideltaDerWFCom}), and using momentum cutoff we get
\begin{eqnarray}\label{PhideltaFirst1}
\Phi _{E}(0,\Lambda)=-\lambda_1 (\Lambda)I_0(E,\Lambda)\left(\frac{A-B}{2\pi}i k+\Phi^{\prime} _{E}(0,\Lambda)\right),\nonumber\\ \Phi^{\prime} _{E}(0,\Lambda)=\lambda_1 (\Lambda) I_2(E,\Lambda)\left(\frac{A+B}{2\pi}+\Phi _{E}(0,\Lambda)\right),
\end{eqnarray}
where
\begin{equation}\label{IE}
    I_k(E,\Lambda)=\frac{1}{2\pi}\int_{-\Lambda}^{\Lambda}\frac{(ip)^k}{E-\sqrt{p^2+m^2}}.
\end{equation}
The divergent part of $I_0(E)$ is similar to the divergent part $I_0(E_B)$. By using eq.(\ref{IEBU}) for the scattering case, we get
\begin{eqnarray}\label{IE0}
I_0(E,\Lambda) &=& \frac{1}{2\pi }
\log\left(\frac{\sqrt{\Lambda^2+m^2}-\Lambda}{\sqrt{\Lambda^2+m^2}+\Lambda}\right)+\frac{E}{\pi \sqrt{E - m^2}} \
\text{arctanh} \left(\frac{\sqrt{E - m^2}}{E}\right).
\end{eqnarray}
Similarly, we get
\begin{eqnarray}\label{IE2}
I_2(E,\Lambda) &=& \frac{1}{2\pi }\left(\Lambda\sqrt{\Lambda^2+m^2}-\left(E^2-\frac{m^2}{2}\right)\log\left(\frac{\sqrt{\Lambda^2+m^2}-\Lambda}{\sqrt{\Lambda^2+m^2}+\Lambda}\right)+2E\Lambda\right)\nonumber\\&+& I_{2c}(E),
\end{eqnarray}
where $I_{2c}(E)$ is the finite part of $I_2(E,\Lambda)$. Again here, it is needless to calculate the value of $I_{2c}(E)$ because it will never play a role in the calculations as we will show this later. From eqs.(\ref{PhideltaFirst1}) we get
\begin{equation}\label{Phidelta1I}
  \Phi _{E}(0,\Lambda)=-\frac{I_0(E,\Lambda)\lambda_1(\Lambda)}{2\pi}\left(\frac{ (A-B)ik+\lambda_1(\Lambda) I_2(E,\Lambda)(A+B)}{1+I_0(E,\Lambda)I_2(E,\Lambda)\lambda_1^2}\right),
\end{equation}
\begin{equation}\label{Phidelta1II}
  \Phi^{\prime} _{E}(0,\Lambda)=\frac{I_2(E,\Lambda)\lambda_1(\Lambda)}{2\pi}\left(\frac{(A+B)-(A-B)ik\lambda_1 I_0(E,\Lambda)}{1+I_0(E,\Lambda)I_2(E,\Lambda)\lambda_1^2(\Lambda)}\right).
\end{equation}
By substituting for the value of $\lambda_1(\Lambda)$ from eq.(\ref{lambdaCut1}) into eq.(\ref{Phidelta1I}) and eq.(\ref{Phidelta1II}), after, we substitute for $I_0(E,\Lambda)$  and $I_2(E,\Lambda)$ from eq.(\ref{IE2}) and eq.(\ref{IE0}), then taking the limit $\Lambda\rightarrow \infty$ we get
\begin{eqnarray}\label{Phi02}
   -\lambda_1 \left(\frac{A-B}{2\pi}ik+\Phi^{\prime} _{E}(0)\right)&=&\frac{A+B}{2\pi(I_{0c}(E_B)-I_{0c}(E))},\ \ \ \ \ \lambda_1=\pm|\lambda_1|,
\end{eqnarray}
\begin{eqnarray}\label{PhiFacOdd}
    \lambda_1(\Lambda) \left(\frac{A+B}{2\pi}+\Phi _{E}(0,\Lambda)\right)\sim \mp \frac{(A+B)\sqrt{\log(2 \Lambda)}}{2 \pi\Lambda(I_{0}(E_B)-I_{0}(E))},\ \ \ \ \ \lambda_1=\pm|\lambda_1|.
\end{eqnarray}
By substituting eq.(\ref{Phi02}) and eq.(\ref{PhiFacOdd}) into eq.(\ref{PhideltaDerWFCom}), we get
\begin{eqnarray}\label{PhideltaDerxFinal}
 \Phi _{E}(x)&=& \frac{(A+B)}{4\pi^2(I_0(E_B)-I_0(E))} \Bigg ( \int_{-\infty }^{\infty }\frac{e^{ipx}dp}{E-\sqrt{p^{2}+m^{2}}}\nonumber\\&\mp&\lim_{\Lambda\rightarrow\infty}\frac{\sqrt{\log(2\Lambda)}}{\Lambda} \int_{-\Lambda }^{\Lambda }\frac{i p e^{ipx}dp}{E-\sqrt{p^{2}+m^{2}}}\Bigg), \hspace{12mm} \lambda_1=\pm |\lambda_1|.
\end{eqnarray}
The odd part of $\Phi _{E}(x)$ term has a factor of $\sqrt{\log(2\Lambda)/\Lambda}$, multiplied by $I_1(x,E,\Lambda)$ which has a pulse at $x=\pm a(E) \Lambda^{-1}$ with a hight $ b(E)\Lambda$ when $\Lambda\rightarrow\infty$, as we explain in Appendix A. On the other hand, the even term diverges logogrammatically at the origin, therefore the odd term can be ignored in the expression of $\Phi_E(x)$ for any $x\in(-\infty,\infty)$. However, this term can not be ignored when taking $\Phi_E^{\prime}(x)$, as we explained in the case of the bound state. The expression $I_0(E_B)-I_0(E)=I_{0c}(E_B)-I_{0c}(E)$ is finite, that is because the divergent terms cancel each other. Our previous non-relativistic treatment suggests that the energy-dependent relativistic running coupling constant  renormalized at the scale $E_B$ is given by
\begin{eqnarray}\label{lambdaDeri}
    \lambda(E,E_B)=\frac{1}{I_{0 c}(E_B)-I_{0 c}(E)}&=&
- \Bigg [\frac{E_B}{2 \pi \sqrt{m^2 - E_B^2}}
\left(\pi + 2 \arcsin\frac{E_B}{m}\right) \nonumber \\
&+&\frac{E}{\pi \sqrt{E^2 - m^2}}
\text{arctanh}\frac{\sqrt{E^2 - m^2}}{E} \Bigg]^{-1}.
\end{eqnarray}
It is easy to prove that for $\Delta E=E-m\ll m$, and $\Delta E_B= E_B-m \ll -m $, the expression of $\lambda(E,E_B)$ reduced to the expression of $\lambda(\Delta E_B)$ in eq.(\ref{LambdaEB}).

Again here, the first integral in eq.(\ref{PhideltaDerxFinal}) can be solved using the contour integration of Figure.2. Accordingly, we get
\begin{eqnarray}\label{PsiXdeltaFirst}
\Psi_E(x)&=& \Bigg[A e^{ikx} +B e^{-ikx}+
\lambda(E,E_B)(A+B)\sqrt{k^2 + m^2} \frac{\sin(k |x|)}{k}
\nonumber \\
&-&\left(\frac{1}{\pi}\lambda(E,E_B)(A+B)\int_m^\infty d\mu
\frac{\sqrt{\mu^2 - m^2}}{\mu^2 + k^2} \exp(- \mu |x|)\right)\nonumber\\&\mp&\lim_{\Lambda\rightarrow\infty}\frac{\sqrt{\log(2\Lambda)}}{\Lambda}(A+B)\lambda(E,E_B) \int_{-\Lambda }^{\Lambda }\frac{ i p \ e^{ipx}dp}{\sqrt{k^2 + m^2}-\sqrt{p^{2}+m^{2}}}\Bigg], \nonumber \\
\lambda_1&=&\pm |\lambda_1|, \ \ \ \ \ \ \ \ \ \ \ E = \sqrt{k^2 + m^2}.
\end{eqnarray}

To understand more the meaning of the wave function in eq.(\ref{PsiXdeltaFirst}), and the constants $A$ and $B$,  we study the reflected and transmitted wave functions for this case. In region I to the left of the contact point, i.e. for $x < 0$, the relativistic reflected wave function takes the following form \cite{Our}
\begin{equation}\label{ReflDel}
\Psi_I(x) = \exp(i k x) + R(k) \exp(- i k x) + C(k) \lambda(E,E_B) \chi_E(x).
\end{equation}
In region II to the right of the contact point, i.e. for $x > 0$, the relativistic transmitted wave function takes the following form
\begin{equation}\label{TransDel}
\Psi_{II}(x) = T(k) \exp(i k x) + C(k) \lambda(E,E_B) \chi_E(x).
\end{equation}
Here, $C(k)$ is a constant that will be determined later, $R(k)$ and $T(k)$ are the reflection and transmission coefficients, and
\begin{equation}
\chi_E(x) = \frac{1}{\pi} \int_m^\infty d\mu \
\frac{\sqrt{\mu^2 - m^2}}{\mu^2 + E^2 - m^2} \exp(- \mu |x|),
\end{equation}
is the branch-cut contribution, which arises in the relativistic case only. This contribution decays exponentially away from the contact point $x = 0$, therefore it has no effect on the scattering wave function at asymptotic distances. By comparing eq.(\ref{PsiXdeltaFirst}) for $x<0$ with eq.(\ref{ReflDel}), and for $x>0$ with eq.(\ref{TransDel}), we get the following relations
\begin{equation}\label{TR}
    T=\frac{k}{k+i\lambda(E,E_B)\sqrt{k^2 + m^2}}, \ \ \ \ \ R=-\frac{i\lambda(E,E_B)\sqrt{k^2 + m^2}}{ k+i\lambda(E,E_B)\sqrt{k^2 + m^2}},
\end{equation}
\begin{eqnarray}
 A=\frac{1}{2}\frac{2k+i\sqrt{k^2 + m^2}}{k+i\sqrt{k^2 + m^2}}, \hspace{10mm} B=\frac{1}{2}R, \hspace{10mm} C(k)=-\frac{1}{\pi}T. \ \ \ \ \ \
\end{eqnarray}

To verify that the resulting system is self-adjoint, we must
examine the orthogonality of the various states. In other words, the scalar product
of the bound state and the scattering states has to vanish, or
\begin{equation}\label{SelfRe1}
    \langle \Psi_B|\Psi_E\rangle=0,
\end{equation}
also the scalar product of two scattering states has to vanish too
\begin{equation}\label{SelfRe2}
    \langle \Psi_{E^{\prime}}|\Psi_E\rangle=0.
\end{equation}
The proof of self-adjointness is explained in appendix B in eq.(\ref{ScalRelB})and eq.(\ref{ScalRelE1}), where we prove that the system is self-adjoint for $\lambda_1=\pm|\lambda_1|$.
\section{ Repulsive and Attractive Scattering States, and the Non-relativistic Limit for the Relativistic Case}

For the relativistic case, and once we remove the cutoff  we have the same bound state for both $\lambda_1=\pm|\lambda_1|$, and the same scattering states for both $\lambda_1=\pm|\lambda_1|$. Moreover, the wave function of the scattering state is like the one for the $\delta$-function potential. To see that, let us take the even part of the wave function in eq.(\ref{PsiXdeltaFirst})
\begin{eqnarray}
\Psi_E(x)+\Psi_E(-x)&=&A(k) \left[\cos(k x) +
\lambda(E,E_B) \frac{\sqrt{k^2 + m^2}}{k} \sin(k |x|) \right.
\nonumber \\
&-&\left. \frac{\lambda(E,E_B)}{\pi} \int_m^\infty d\mu
\frac{\sqrt{\mu^2 - m^2}}{\mu^2 + k^2} \exp(- \mu |x|)\right], \
\end{eqnarray}
This exactly the same expression of the scattering wave function of the $\delta$-function potential that was derived in \cite{Our}. The same goes for the bound state.

From eq.(\ref{lambdaDeri}) and eq.(\ref{IEBc0}), bound and strong bound states ($|E_B|<m$) are correspond to attractive $\delta'$-function potential, because then $ \lambda(E,E_B)<0$ . On the other hand, for ultra- strong bound state ($E_B < - m$ ), the value of $I_{0c}(E_B)$ is given by eq.(\ref{IEBU}), and therefore it gives $ \lambda(E,E_B)>0$ for $E>E_B$, (see Figure 5). This correspond to  a repulsive $\delta'$-function potential.
\begin{figure}[t]
\begin{center}
\epsfig{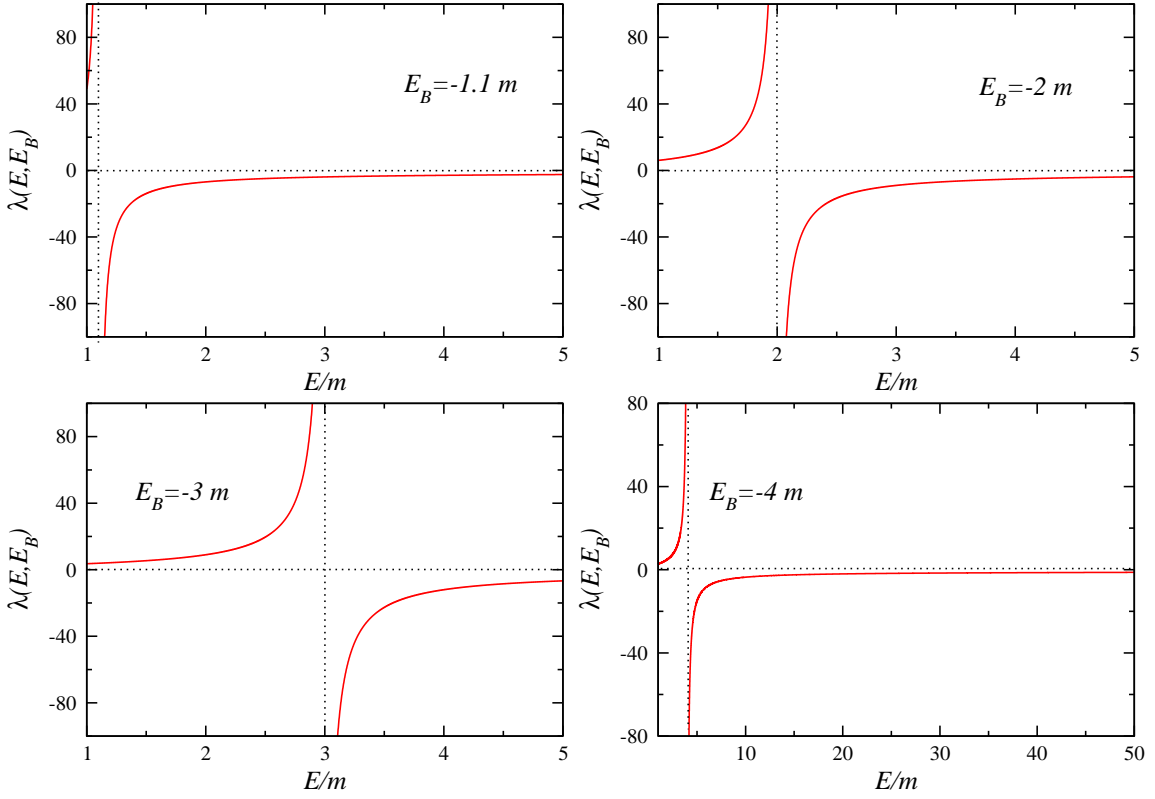}
\end{center}
\caption{\it The running coupling $\lambda(E,E_B)$ as a function of the scattering energy $E$ in the units of $m$, for $E_B=-1.1m, -2m, -3m,$ and $-4m$. The graph in the lower right corner was extended to large values of $E$ in order to illustrate the asymptotic freedom of the system when $\lambda(E,E_B)\rightarrow 0$ as $E\rightarrow \infty$}
\label{isofig3}
\end{figure}

By taking $\varkappa/m \rightarrow 0$, we get the non-relativistic limit for the relativistic bound state. Accordingly, eq.(\ref{BRState}) gives
\begin{eqnarray}\label{NRbou}
\Psi_B(x) &=&\sqrt{\varkappa}
\Big[\frac{\varkappa}{m \pi}  \int_m^\infty d\mu
\frac{\sqrt{\mu^2 - m^2}}{\mu^2 - \varkappa^2} \exp(- \mu |x|) +
\exp(- \varkappa |x|)\Big]\nonumber\\ &\pm&\sqrt{\varkappa}\lim_{\Lambda\rightarrow\infty}\frac{2m\sqrt{\log(2\Lambda)}}{\Lambda}(A+B)\lambda(E,E_B) \int_{-\Lambda }^{\Lambda }\frac{ i p \ e^{ipx}dp}{2\Delta E_B m-p^2}.
\end{eqnarray}
This means that the wave function
reduces to the bound state for the non-relativistic case in eq.(\ref{NonRelWFfinal1}). However, the divergence at the origin of the relativistic wave function persists for any
non-zero value of $\varkappa/m$. Moreover, the last integral in eq.(\ref{NRbou}) does not reduce to the analogous form in eq.(\ref{NonRelWFfinal1}), however, this term can be ignored if we remove the cutoff. The non-relativistic limit for the relativistic scattering states is
\begin{eqnarray}\label{PsiXdeltaFirstNonLim}
\Psi_E(x)&=& A e^{ikx} +B e^{-ikx}+
(A+B)\lambda(E_B) \frac{m \sin(k |x|)}{k} \nonumber \\
&-&\left(\frac{1}{\pi}\lambda(E_B)(A+B)\int_m^\infty d\mu
\frac{\sqrt{\mu^2 - m^2}}{\mu^2 + k^2} \exp(- \mu |x|)\right)\nonumber\\&\mp&\lim_{\Lambda\rightarrow\infty}\frac{2m\sqrt{\log(2\Lambda)}}{\Lambda}(A+B)\lambda(E_B) \int_{-\Lambda }^{\Lambda }\frac{ i p \ e^{ipx}dp}{k^2-p^2}\Bigg], \nonumber \\ 
\end{eqnarray}
where $E=k^2/ 2m$. The same argument goes for the non-relativistic limit of the scattering states.
\par If we take the limit $E_B \rightarrow - \infty$, the running coupling constant in eq.(\ref{lambdaDeri}) takes the form
\begin{equation}
\lambda(E,E_B) \rightarrow - \left[\frac{E}{\pi \sqrt{E^2 - m^2}} \
\text{arctanh}\frac{\sqrt{E^2 - m^2}}{E} -
\frac{1}{\pi} \log\left(\frac{- 2 E_B}{m}\right)\right]^{-1}.
\end{equation}
For small non-relativistic energies $\Delta E = E - m \ll m$, this reduces to
\begin{equation}\label{LambdaEBg}
\lambda \rightarrow \frac{\pi}{\log(- 2 E_B/m)} > 0.
\end{equation}
Therefore we are reaching the non-relativistic limit for a repulsive $\delta'$-function potential with a coupling parameter $\lambda(E_B)>0$. This exactly the same as the case of the $\delta$- function potential \cite{Our}. An  important feature of the non-relativistic case is that it has only an attractive $\delta'$-function potential, a result that has been also reached by \cite{Widmer}. In contrast, the non-relativistic limit of the relativistic case for  ultra-strong bound state gives a repulsive $\delta'$-function potential with $\lambda(E_B)>0$ in eq.(\ref{PsiXdeltaFirstNonLim}). At first glance, this seems to be a paradox. However, the fact that contact interactions happen at very short distances can explain the issue. Very short distances mean high momentum transfer, therefore even for non-relativistic limit energies, the particle still influenced by the powers of $p$ higher than two in the expansion of the pseudo-differential operator.

\section{Conclusions}
The investigation of the $\delta'$- function potential  in 1-dimensional non-relativistic and relativistic quantum mechanics require regularization. On the other hand,  renormalizing the coupling constant $\lambda_1$ using renormalization methods in the  usual sense is not possible due to the square root in eq.(\ref{labdaNonCut1}) and eq.(\ref{lambdaCut1}). However, $\lambda_1$ always appears in the equation of motion as $\lambda_1 \Psi_B(0)$, and $\lambda_1 \Psi'_B(0)$, and therefore this suggests to redefine the concept of renormalization in this case to a renormalized $\lambda_1 \Psi_B(0)$, and $\lambda_1 \Psi_B'(0)$. This could be done by proving that they are  finite quantities under the regularization scheme. The cutoff regularization is successful in calculating the order of smallness of $\lambda_1 $, $\Psi_B(0)$, and $\Psi'_B(0)$. When removing the cutoff we get  $\lambda_1 \Psi_B(0)\rightarrow 0$, while the value of $\lambda_1 \Psi_B'(0)$ is obtained from the normalization condition. Having these information we are able to obtain the wave function for the bound state, and scattering states in both relativistic and non-relativistic cases, and  without the the need for the explicit form of $\lambda_1$. Before removing the cutoff, in both of the non-relativistic and relativistic cases, there are 2-parameters family of self-adjoint extensions $\lambda_1=\pm|\lambda_1|$. However, once we remove the cutoff, we end up only with one parameter, that is the coupling constant $\lambda(\Delta E_B)$ in the non-relativistic case, and  $\lambda(E,E_B)$ energy-dependent relativistic running coupling constant in the relativistic case.

After removing the cutoff, the resultant wave functions, bound and scattering states have exactly the same expression of the analogous ones in the relativistic $\delta$-function potential case. As a result, we have the same interesting features of the $\delta$-function potential like, asymptotic freedom, dimensional transmutation, and an infra-red conformal fixed point in the massless limit that was discussed in our previous paper \cite{Our}. The only difference is that $\lambda_1 \Psi_B'(0)$ is a non-zero constat while  $\lambda_1 \Psi_B(0)\rightarrow 0$ in the case of the $\delta'$-function potential, while $\lambda_1 \Psi_B(0)$ is non-zero constant in the case of the $\delta$-function potential. However this will not affect the probability density in both problems, because in the two problems, both of $\lambda_1 \Psi'_B(0)$ and $\lambda_1 \Psi_B(0)$ have the same dependency on $E_B$, as it is clear from eq.(\ref{NormCon}) and eq.(\ref{Psi0PPValueF}). It is not clear if the work in \cite{Alb97} leads to the same conclusion.

One of the important results of this work is highlighting the fact that the non-relativistic limit of the relativistic case does not lead exactly to the non-relativistic solution. The non-relativistic case  has only an attractive $\delta'$-function potential. In contrast, the non-relativistic limit of the relativistic case, and for ultra-strong bound state gives a repulsive $\delta'$-function potential, where $\lambda(E_B)>0$. This is explained by the notion  that $\delta'$-function potential is a contact interaction that takes place at  very short distances, which mean high momentum transfer. Therefore, even for non-relativistic limit energies, the particle still influenced by powers of $p$ higher than two in the expansion of the pseudo-differential operator. This  also explain why the divergence at the origin persist when taking the non-relativistic limit of the relativistic case. The $\delta'$-function potential reveal this issue more than the $\delta$-function potential, because in the the $\delta$-function potential, we do have a repulsive solution for the non-relativistic case.

The ideas and procedures discussed in this paper can be useful in solving the relativistic, and non-relativistic Schr\"{o}dinger equation for potentials with higher derivatives of the delta function. We can also investigate higher dimensions, and check the self-adjointness of such systems using the procedure explained in appendix B. It is well known that some regularization methods give rise to a nonself-adjoint Hamiltonian  \cite{Phi98}, and It would be interesting to verify this independently. In  future work, there  is a possibility  of a successful investigation  of a contact interaction for a two-particles system using this approach. Such system has a total energy $E=\sqrt{P^2+M^2}$, where $P$,  and $M$ are the total momentum, and the rest-energy of the system respectively. In this case, we have to construct a boost operator, and prove that the Poincar\'{e} algebra is respected. The  mass spectrum for this case offers an interesting result that could be matched with a mass spectrum from a quantum field theory.
\section*{Appendix A: The Order of Smallness of the Odd Part of the Relativistic Wave Function}
The wave function in eq.(\ref{PsideltaDerWFCom}) can be written as
\begin{equation}
    \Psi_B(x,\Lambda)=\Upsilon_{Be}(x,\Lambda)+\Upsilon_{Bo}(x,\Lambda),
\end{equation}
where $\Upsilon_{Be}(x,\Lambda)$ and  $\Upsilon_{Bo}(x,\Lambda)$ is the even and odd part of the relativistic bound state respectively. Accordingly,  $\Upsilon_{Bo}(x,\Lambda)$ can be written as
\begin{eqnarray}\label{Ups}
 \Upsilon_{Bo}(x,\Lambda)&=& \pm C_1\sqrt{-\frac{I_{0}(E_B,\Lambda)}{I_{2}(E_B,\Lambda)}}I_{1}(x,E_B,\Lambda)\nonumber\\&=& \pm C_1 \sqrt{-\frac{I_{0}(E_B,\Lambda)}{I_{2}(E_B,\Lambda)}} \int_{-\Lambda}^{\Lambda}\frac{i p \exp(ip x)}{E_B-\sqrt{p^2+m^2}} dp \nonumber\\ &=&\pm C_1 \sqrt{-\frac{I_{0}(E_B,\Lambda)}{I_{2}(E_B,\Lambda)}}\sum_{k=1}^{\infty} \int_{-\Lambda}^{\Lambda}\frac{(i p)^k x^{k-1}}{(E_B-\sqrt{p^2+m^2})(k-1)!} dp.
\end{eqnarray}
For arbitrary $x$, the value of $\Upsilon_{Bo}(x,\Lambda)$ in  eq.(\ref{Ups}) vanishes as $\Lambda\rightarrow\infty$. That is because $\sqrt{-I_{0}(E_B,\Lambda)/I_{2}(E_B,\Lambda)}$ goes to zero as $\Lambda\rightarrow\infty$. However, this is not correct for all values of $x$, there is a spike in the value of $\Upsilon_{Bo}(x,\Lambda)$ in the neighborhood of  $x=\pm \varsigma$. The numerical calculations show that the extrema values of $I_{1}(x,E_B,\Lambda)$ is proportional to $\Lambda$, and the value of $ \varsigma$ is inversely proportional to $\Lambda$, as $\Lambda\rightarrow\infty$. For large enough $\Lambda$, the series in eq.(\ref{Ups}) converges. The integral in the right hand side of eq.(\ref{Ups}) can be obtained analytically for all values of $k$. Accordingly, we can write the above equations as
\begin{eqnarray}\label{Ups1}
   \Upsilon_{Bo}(x,\Lambda)&=& \pm C_1\sqrt{-\frac{I_{0}(E_B,\Lambda)}{I_{2}(E_B,\Lambda)}} \nonumber \\ &\times& \sum_{k=1}^{\infty}2\Lambda^{2k+1}\Bigg(\frac{m F_1(\frac{1}{2}+k;-\frac{1}{2},1; \frac{3}{2}+k; \frac{-\Lambda^2}{m^2},-\frac{\Lambda^2}{m^2-E_B^2})}{(m^2-E_B^2)(2k+1)(2k-1)!)}\nonumber \\&+&
   \frac{E_B \ _{2}F_1(1,\frac{1}{2}+k; \frac{3}{2}+k; -\frac{\Lambda^2}{m^2-E_B^2})}{(m^2-E_B^2)(2k+1)(2k-1)!}\Bigg)(-1)^{k-1}\varsigma^{2k-1},
\end{eqnarray}
where $F_1(a_1;a_2,a_3; a_4;z_1,z_2)$ is the Appel hypergeometric function with two variables, and $_{2}F_1(a_1,a_2; a_3; z)$ is a hypergeometric function with one variable. The value of $\varsigma$ can be obtained from the first maximum of $I_{1}(x,E_B,\Lambda)$, or
\begin{equation}
   \frac{\partial I_{1}(x,E_B,\Lambda)}{\partial x}|_{x=\varsigma}=\int_{-\Lambda}^{\Lambda}\frac{- p^2 \exp(ip x)}{(E_B-\sqrt{p^2+m^2})^2} dp|_{x=\varsigma}=0.
\end{equation}
From the above equation, and using  eq.(\ref{Ups1}), the value of $\varsigma$ can be calculated by solving numerically the following equation  
\begin{eqnarray}\label{Ups2}
   \sum_{k=1}^{\infty}2\Lambda^{2k+1}\Bigg(\frac{m F_1(\frac{1}{2}+n;-\frac{1}{2},1; \frac{3}{2}+n; \frac{-\Lambda^2}{m^2},-\frac{\Lambda^2}{m^2-E_B^2})}{(m^2-E_B^2)(2k+1)(2k-2)!}\nonumber \\+
   \frac{E_B \ _{2}F_1(1,\frac{1}{2}+n; \frac{3}{2}+n; -\frac{\Lambda^2}{m^2-E_B^2})}{(m^2-E_B^2)(2k+1)(2k-2)!}\Bigg)(-1)^{k-1}\varsigma^{2k-2}=0.
\end{eqnarray}
There are infinite number of values for $\varsigma$ that satisfie the above equation. We are only interested in the smallest $\varsigma$. As $\Lambda\rightarrow\infty$, it can be proved that $\varsigma\rightarrow a(E_B)\Lambda^{-1}$. By substituting the obtained value of $\varsigma$ in eq.(\ref{Ups1}) we get
\begin{eqnarray}\label{Ups3}
 \Upsilon_{Bo}(\pm \varsigma,\Lambda)&=&\pm C_1\sqrt{-\frac{I_{0}(E_B,\Lambda)}{I_{2}(E_B,\Lambda)}}b(E_B) \Lambda, \ \ \ \ \ \ \ \lambda_1=\pm |\lambda_1|
\end{eqnarray}

For the scattering states, we write
\begin{equation}
    \Phi_E(x,\Lambda)=\Upsilon_{Ee}(x,\Lambda)+\Upsilon_{Eo}(x,\Lambda).
\end{equation}
where $\Upsilon_{Ee}(x,\Lambda)$ is the even part, and  $\Upsilon_{Eo}(x,\Lambda)$ is the odd part of $\Phi_E(x,\Lambda)$. From eq.(\ref{PhideltaDerxFinal}), the odd part takes the following form
 \begin{equation}
   \Upsilon_{Eo}(x,\Lambda)= \mp\frac{\sqrt{\log(2 \Lambda)}}{\Lambda}\int_{-\Lambda}^{\Lambda}\frac{i p \exp(ip x)}{E-\sqrt{p^2+m^2}} dp, \ \ \ \ \ \ \ \lambda_1=\pm |\lambda_1|
 \end{equation}
 The same mathematical treatment of the bound state can be repeated for scattering states, mainly by replacing $E_B$ with $E$ in eq.(\ref{Ups}), eq.(\ref{Ups1}) and eq.(\ref{Ups2}). In this case, we also find numerically that when $\Lambda\rightarrow\infty$, the extrema of $I_{1}(x,E,\Lambda)$ are at $x=\varsigma\rightarrow a(E_B)\Lambda^{-1}$, and
 \begin{equation}
   \Upsilon_{Eo}(\pm\varsigma,\Lambda)= \mp\frac{\sqrt{\log(2 \Lambda)}}{\Lambda}b(E)\Lambda, \ \ \ \ \ \ \ \lambda_1=\pm |\lambda_1|.
 \end{equation}
\section*{Appendix B: Self-Adjointness of the System}
\subsection*{The non-relativistic case}
The scalar product of $\Psi_B(x)$ and $\Psi_E(x)$ is
\begin{eqnarray}\label{OSBAp}
\langle \Psi_B|\Psi_E\rangle&=&\lim_{\Lambda\rightarrow\infty}
\frac{C_1}{ \pi} \int_{-\Lambda}^{\Lambda} dp  \frac{1}{2m \Delta E_B - p^2}\left(1\pm ip \sqrt{-\frac{I_{0}(\Delta E_B,\Lambda)}{I_{2}(\Delta E_B,\Lambda)}}\right)\nonumber \\
&\times&[A\delta(p - \sqrt{2m\Delta E}) +B \delta(p + \sqrt{2m\Delta E}) +
\widetilde \Phi_{E}(p,\Lambda)],
\end{eqnarray}
where
\begin{equation}\label{PhiAp}
\tilde{\Phi}_{E}(p,\Lambda)=\frac{2m\lambda_1(\Lambda)}{2mE-p^2}\left( ip\left( \frac{A+B}{2\pi }+\Phi_{E}(0,\Lambda)\right)-\left( \left( \frac{A-B}{2\pi }\right) ik+\Phi_{ E}^{\prime}(0,\Lambda)\right)\right).
\end{equation}
From the above equation and eq.(\ref{OSBAp}) we get
\begin{eqnarray}
\langle \Psi_B|\Psi_E\rangle&\sim&\frac{A+B}{2 m(\Delta E_B -\Delta E)}+\lim_{\Lambda\rightarrow\infty}\Bigg( \pm \sqrt{2m\Delta E}\frac{A-B}{2 m(\Delta E_B - \Delta E)}\sqrt{-\frac{I_{0}(E_B,\Lambda)}{I_{2}(\Delta E_B,\Lambda)}}\nonumber \\
&+&\frac{(A+B)}{2\pi I_0(\Delta E_B) } 
\int_{-\Lambda}^{\Lambda} dp \ \frac{\left(1\pm ip \sqrt{-\frac{I_{0}(\Delta E_B,\Lambda)}{I_{2}(\Delta E_B,\Lambda)}}\right)}{2m\Delta E_B - p^2} \ \frac{1}{2m\Delta E - p^2}
\nonumber \\ &\pm& (A+B)\sqrt{\frac{m}{-2\pi I_0(\Delta E_B,\Lambda)}}\sqrt{\frac{1}{\Lambda}}\int_{-\Lambda}^{\Lambda} dp \ \Bigg(\frac{\left(1\pm ip \sqrt{-\frac{I_{0}(\Delta E_B,\Lambda)}{I_{2}(\Delta E_B,\Lambda)}}\right)}{2m\Delta E_B - p^2 }\nonumber\\ &\times & \frac{ip}{2m\Delta E - p^2}\Bigg) \Bigg).
\end{eqnarray}
The integrals in the above equation can be obtained by using the following relation
\begin{equation}\label{IEIBRe}
    \frac{m}{\pi}\int_{-\Lambda}^{\Lambda} dp\frac{(ip)^{k}}{2m \Delta E_B - p^2}\frac{1}{2m \Delta E - p^2}=\frac{I_k(\Delta E_B,\Lambda)-I_k(\Delta E,\Lambda)}{\Delta E-\Delta E_B}.
\end{equation}

The above relation means that any integral involves odd powers of $p$ vanishes. For $\Lambda\rightarrow \infty$, the term involves $p^2$ gives 
\begin{eqnarray}
    \frac{m}{\pi}\int_{-\Lambda}^{\Lambda} dp\frac{-p^2}{2m \Delta E_B - p^2}\frac{1}{2m \Delta E - p^2}&=&\frac{I_2(\Delta E_B,\Lambda)-I_2(\Delta E,\Lambda)}{\Delta E-\Delta E_B}\nonumber \\
    &\sim& \sqrt{\Delta E_B}+i \sqrt{\Delta E},
\end{eqnarray}
This means that after taking the limit $\Lambda\rightarrow\infty$, the only terms left in eq.(\ref{OSBAp}) are
\begin{equation}\label{ScalRelB}
    \langle \Psi_B|\Psi_E\rangle \sim \frac{A+B}{2m(\Delta E_B -\Delta E)} +\frac{A+B}{2m(\Delta E -\Delta E_B)}=0 \ \ \ \ \ \ \ \ \  \lambda_1=\pm |\lambda_1| 
\end{equation}
The scalar product of $\Psi_{E'}(x)$ and $\Psi_E(x)$ is
\begin{eqnarray}
\label{OSBAp2}
\langle \Psi_{E^{\prime}}|\Psi_E\rangle&=&\lim_{\Lambda\rightarrow\infty}
 \int_{-\Lambda}^{\Lambda} \frac{dp}{2 \pi}
[A^{*\prime}\delta(p - \sqrt{2 m\Delta E'}) +B^{*\prime} \delta(p + \sqrt{2 m \Delta E'}) +
\widetilde \Phi^{*}_{E'}(p,\Lambda)]\nonumber  \\ &\times&[A\delta(p - \sqrt{2 m \Delta E}) +B \delta(p + \sqrt{2 m \Delta E}) +
\widetilde \Phi_E(p,\Lambda)]
\end{eqnarray}
Using eq.(\ref{PhiAp}), we can write eq.(\ref{OSBAp2}) as
\begin{eqnarray}\label{EEpNon}
\langle \Psi_{E'}|\Psi_E\rangle&\sim&
 (A^{*\prime}A+B^{*\prime}B)\delta(\sqrt{2m \Delta E} - \sqrt{2m\Delta E'})\nonumber\\ &+&\frac{(A+B)(A^{*\prime}+B^{*\prime})\lambda(\Delta E_B)}{4\pi m(\Delta E- \Delta E')}+\frac{(A+B)(A^{*\prime}+B^{*\prime})\lambda(\Delta E_B)}{4\pi m(\Delta E'-\Delta E)}
\nonumber\\\lim_{\Lambda\rightarrow\infty}\Bigg(&\pm &\sqrt{\frac{-\lambda(\Delta E_B)m}{2\pi}} \sqrt{\frac{1}{\Lambda}}\frac{(A+B)(A^{*\prime}-B^{*\prime})}{2m \Delta E-2m \Delta E'}i\sqrt{2m\Delta E'}
\nonumber\\&\pm&\sqrt{\frac{-\lambda(\Delta E_B)m}{2\pi}} \sqrt{\frac{1}{\Lambda}}\frac{(A-B)(A^{*\prime}+B^{*\prime})}{2m \Delta E'-2m\Delta  E}i\sqrt{2m \Delta E}
\nonumber\\&+&(A+B)(A^{*\prime}+B^{*\prime})\int_{-\Lambda}^{\Lambda} dp\Big(\frac{\frac{\lambda(\Delta E_B)m}{\pi}+ip \sqrt{\frac{-\lambda(\Delta E_B)m}{2\pi}} \sqrt{\frac{1}{\Lambda}}}{2m\Delta E -p^2}
\nonumber\\ &\times &\frac{\frac{\lambda(\Delta E_B)m}{\pi}-ip \sqrt{\frac{-\lambda(\Delta E_B)m}{2\pi}} \sqrt{\frac{1}{\Lambda}}}{2m\Delta E' -p^2}\Big)\Bigg).
\end{eqnarray}
We can use  eq.(\ref{IEIBRe}) in the above equation by replacing $\Delta E_B$ with $\Delta E'$, then any integral involves power one of $p$ vanishes. Moreover, we get
\begin{eqnarray}
    \frac{m}{\pi}\int_{-\Lambda}^{\Lambda} dp\frac{-p^2}{2m \Delta E' - p^2}\frac{1}{2m \Delta E - p^2}&=&\frac{I_2(\Delta E',\Lambda)-I_2(\Delta E,\Lambda)}{\Delta E-\Delta E'}\nonumber \\
    &\sim& \sqrt{\Delta E'}- \sqrt{\Delta E}.
\end{eqnarray}
Accordingly, after taking the limit $\Lambda\rightarrow \infty$, we can write eq.(\ref{EEpNon}) as
\begin{eqnarray}
\langle \Psi_{E'}|\Psi_E\rangle&\sim&
\delta(\sqrt{2m \Delta E} - \sqrt{2m E'})=\delta(k - k')\ \ \ \ \  \ \   \lambda_1=\pm |\lambda_1|.
\end{eqnarray}
\subsection{The relativistic case}
The scalar product of $\Psi_B(x)$ and $\Psi_E(x)$ is
\begin{eqnarray}
\label{OSBRelAp1}
\langle \Psi_B|\Psi_E\rangle&=&\lim_{\Lambda\rightarrow\infty}
\frac{C_1}{2 \pi} \int_{-\Lambda}^{\Lambda} dp  \frac{1}{E_B - \sqrt{p^2 + m^2}}\left(1\pm ip \sqrt{-\frac{I_{0}(E_B,\Lambda)}{I_{2}(E_B,\Lambda)}}\right)\nonumber \\
&\times&[A\delta(p - \sqrt{E^2 - m^2}) +B \delta(p + \sqrt{E^2 - m^2}) +
\widetilde \Phi_E(p,\Lambda)],
\end{eqnarray}
where
\begin{equation}\label{PhiReAp}
\tilde{\Phi}_E(p,\Lambda)=\frac{\lambda_1(\Lambda)}{E-\sqrt{p^{2}+m^{2}}}\left( ip\left( \frac{A+B}{2\pi }+\Phi _{E}(0,\Lambda)\right)
-\left( \left( \frac{A-B}{2\pi }\right) ik+\Phi _{E}^{\prime }(0,\Lambda)\right)\right).
\end{equation}
The above two equations together with eq.(\ref{Phi02}) and eq.(\ref{PhiFacOdd}) give
\begin{eqnarray}\label{ScalReAp}
\langle \Psi_B|\Psi_E\rangle&\sim&\frac{A+B}{E_B - E}+ \lim_{\Lambda\rightarrow\infty}\Bigg(\pm i\sqrt{E^2 - m^2}\frac{A-B}{E_B - E}\sqrt{-\frac{I_{0}(E_B,\Lambda)}{I_{2}(E_B,\Lambda)}}\nonumber \\
&\pm&\frac{ (A+B)}{2\pi(I_0(E_B) - I_0(E))}
\int_{-\Lambda}^{\Lambda} dp \ \frac{\left(1\pm ip \sqrt{-\frac{I_{0}(E_B,\Lambda)}{I_{2}(E_B,\Lambda)}}\right)}{E_B - \sqrt{p^2 + m^2}} \ \frac{1}{E - \sqrt{p^2 + m^2}}
\nonumber \\ &\mp& \frac{(A+B)\sqrt{\log(2 \Lambda)}}{2\pi\Lambda(I_{0}(E_B)-I_{0}(E))}\int_{-\Lambda}^{\Lambda} dp \ \frac{\left(1\pm ip \sqrt{-\frac{I_{0}(E_B,\Lambda)}{I_{2}(E_B,\Lambda)}}\right)}{E_B - \sqrt{p^2 + m^2} }\ \frac{ip}{E - \sqrt{p^2 + m^2}}\Bigg) \nonumber\\
\end{eqnarray}
The integrals in the above equation can be obtained by using the following relation
\begin{equation}\label{IEIBreR}
    \frac{1}{2\pi}\int_{-\Lambda}^{\Lambda} dp\frac{(ip)^{k}}{E_B - \sqrt{p^2 + m^2}}\frac{1}{E - \sqrt{p^2 + m^2}}=\frac{I_k(E_B,\Lambda)-I_k(E,\Lambda)}{E-E_B}
\end{equation}
The above relation means that any integral involves odd power of $p$ vanishes. For $\Lambda\rightarrow \infty$, the term involves $p^2$ gives 
\begin{equation}
    \frac{1}{2\pi}\int_{-\Lambda}^{\Lambda} dp\frac{-p^2}{E_B - \sqrt{p^2 + m^2}}\frac{1}{E - \sqrt{p^2 + m^2}}=\frac{I_2(E_B,\Lambda)-I_2(E,\Lambda)}{E-E_B}\sim \Lambda.
\end{equation}
From the above two equations, and  after taking  the limit $\Lambda\rightarrow\infty$, we get
\begin{equation}
    \langle \Psi_B|\Psi_E\rangle =0, \ \ \ \ \ \ \ \ \lambda_1=\pm|\lambda_1|.
\end{equation}

The scalar product of $\Psi_{E'}(x)$ and $\Psi_E(x)$ is
\begin{eqnarray}
\label{ScalReEEp}
\langle \Psi_{E^{\prime}}|\Psi_E\rangle&=&\lim_{\Lambda\rightarrow\infty}
 \int_{-\Lambda}^{\Lambda} \frac{dp}{2\pi}
[A^{*\prime}\delta(p - \sqrt{E'^2 - m^2}) +B^{*\prime} \delta(p + \sqrt{E'^2 - m^2}) +
\widetilde \Phi_{E'}(p,\Lambda)]\nonumber  \\ &\times&[A\delta(p - \sqrt{E^2 - m^2}) +B \delta(p + \sqrt{E^2 - m^2}) +
\widetilde \Phi_E(p,\Lambda)].
\end{eqnarray}
Using eq.(\ref{PhiReAp}) we can write eq.(\ref{ScalReEEp}) as
\begin{eqnarray}\label{ScalReEEp2}
\langle \Psi_{E'}|\Psi_E\rangle&\sim&
(A^{*\prime}A+B^{*\prime}B)\delta(\sqrt{E^2 - m^2} - \sqrt{{E'}^2 - m^2})\nonumber\\ &+&\frac{(A+B)(A^{*\prime}+B^{*\prime})\lambda(E,E_B)}{2\pi(E-E')}+\frac{(A+B)(A^{*\prime}+B^{*\prime})\lambda(E',E_B)}{2\pi(E'-E)}
\nonumber\\\lim_{\Lambda\rightarrow\infty}\Bigg(&\mp&\frac{\sqrt{\log(2 \Lambda)}}{\Lambda}\frac{(A+B)(A^{*\prime}-B^{*\prime})\lambda(E,E_B)}{2\pi(E-E')}i\sqrt{E'^2-m^2}
\nonumber\\&\mp&\frac{\sqrt{\log(2 \Lambda)}}{\Lambda}\frac{(A-B)(A^{*\prime}+B^{*\prime})\lambda(E',E_B)}{2\pi(E'-E)}i\sqrt{E^2-m^2}
\nonumber\\&+&\lambda(E',E_B)\lambda(E,E_B)(A+B)(A^{*\prime}+B^{*\prime})\int_{-\Lambda}^{\Lambda} \frac{dp}{4\pi^2}\Big(\frac{1\mp ip \frac{\sqrt{\log(2 \Lambda)}}{\Lambda}}{E - \sqrt{p^2 + m^2}}
\nonumber\\ &\times &\frac{1\mp ip \frac{\sqrt{\log(2\Lambda)}}{\Lambda}}{E' - \sqrt{p^2 + m^2}}\Big)\Bigg).
\end{eqnarray}
We can use  eq.(\ref{IEIBreR}) in the above equation by replacing $E_B$ with $E'$, then any integral involves power one of $p$ vanishes. I addition, we get
\begin{equation}
    \frac{1}{2\pi}\int_{-\Lambda}^{\Lambda} dp\frac{-p^2}{E' - \sqrt{p^2 + m^2}}\frac{1}{E - \sqrt{p^2 + m^2}}=\frac{I_2(E',\Lambda)-I_2(E,\Lambda)}{E-E'}\sim \Lambda.
\end{equation}
Accordingly we can write eq.(\ref{ScalReEEp2}) as
\begin{equation}\label{ScalRelE1}
    \langle \Psi_E'|\Psi_E\rangle\sim \delta(\sqrt{E^2 - m^2} - \sqrt{{E'}^2 - m^2}), \ \ \ \ \ \ \ \ \ \ \lambda_1=\pm|\lambda_1|.
\end{equation}

\section*{Acknowledgments}
This publication was made possible by the NPRP grant \# NPRP 5 - 261-1-054 from
the Qatar National Research Fund (a member of the Qatar Foundation). The
statements made herein are solely the responsibility of the authors. \text


\begin{thebibliography}{34}
\bibitem{Bak73}
F.\ Bakke and H.\ Wergeland, Physica 69 (1973) 5.

\bibitem{Alm84}
C.\ Almeida and A.\ Jabs, Am.\ J.\ Phys.\ 52 (1984) 921.

\bibitem{Str06}
F.\ W.\ Strauch, Phys.\ Rev.\ A73 (2006) 069908.

\bibitem{AlH09}
M.\ Al-Hashimi and U.-J.\ Wiese, Ann.\ Phys.\ 324 (2009) 2599.

\bibitem{Cal88}
D.\ J.\ E.\ Callaway, Phys.\ Rep.\ 167 (1988) 241.

\bibitem{Tho79}
C.\ Thorn, Phys.\ Rev.\ D19 (1979) 639.

\bibitem{Beg85}
M.\ A.\ B.\ Beg and R.\ C.\ Furlong, Phys.\ Rev.\ D31 (1985) 1370.

\bibitem{Hag90}
C.\ R.\ Hagen, Phys.\ Rev.\ Lett.\ 64 (1990) 503.

\bibitem{Jac91}
R.\ Jackiw, M.\ A.\ B.\ Beg Memorial Volume, A.\ Ali and P.\ Hoodbhoy, Eds.,
World Scientific, Singapore (1991).

\bibitem{Fer91}
J.\ Fernando Perez and F.\ A.\ B.\ Coutinho, Am.\ J.\ Phys.\ 59 (1991) 52.

\bibitem{Gos91}
P.\ Gosdzinsky and R.\ Tarrach, Am.\ J.\ Phys.\ 59 (1991) 70.

\bibitem{Mea91}
L.\ R.\ Mead and J.\ Godines, Am.\ J.\ Phys.\ 59 (1991) 935.

\bibitem{Man93}
C.\ Manuel and R.\ Tarrach, Phys.\ Lett.\ B328 (1994) 113.

\bibitem{Phi98}
D.\ R.\ Phillips, S.\ R.\ Beane, and T.\ D.\ Cohen, Ann.\ Phys.\ 263 (1998)
255.

\bibitem{Bol72}
C.\ G.\ Bollini and J.\ J.\ Giambiagi, Nuovo Cim.\ 12B (1972) 20.

\bibitem{Bol72a}
C.\ G.\ Bollini and J.\ J.\ Giambiagi, Phys.\ Lett.\ 40B (1972) 566.

\bibitem{tHo72}
G.\ t' Hooft and M.\ Veltman, Nucl.\ Phys.\ B44 (1972) 189.

\bibitem{Bie14}
W.\ Bietenholz and L.\ Prado, Physics Today 67 (2014) 38.

\bibitem{Alb97}
S.\ Albeverio and P.\ Kurasov, Lett.\ Math.\ Phys.\ 41 (1997) 79.
\bibitem{Our}
M.\ H.\ Al-Hashimi, A.\ Shalaby, U.-J.\ Wiese, Phys.\ Rev.\ D 89 (2014) 125023.
\bibitem{Our2}
M.\ H.\ Al-Hashimi, A.\ Shalaby, arXiv:1406.3265.
\bibitem{Hold}
F.\ Gesztesy, H.\ Holden   J.\ Phys.\ A: \ Math.\ Gen.\ 20 (1987)5157.

\bibitem{Albbook}
S.\ Albeverio, F.\ Gesztesy, R.\ Hoeg-Krohn, and H.\ Holden, Solvable Models
in Quantum Mechanics, Texts and Monographs, Springer, New York (1988).

\bibitem{Arnbak1}
P.\ Christiansen, H.\ Arnbak, A.\ Zolotaryuk, V.\ Ermakov, Y.\ Gaididei, J.\ Phys.\ A: \ Math.\ Gen.\ 36 (2003) 7589.
\bibitem{Pes}
M.\ E.\ Peskin, and D.\ V.\ Schroeder, An Introduction to Qunantum  Field Theory, Addison-Wesley Advanced Book Program (1995).
\bibitem{Zhao}
B.\ Zhao, J.\ Phys.\ A:\ Math.\ Gen.\ 25 (1992) L617.

\bibitem{Griff}
D.\ Griffiths, J.\ Phys.\ A:\ Math.\ Gen.\ 26 (1993) 2265.

\bibitem{Toy}
F.\ Toyama1, Y.\ Nogami, J.\ Phys.\ A:\ Math.\ Theor.\ 40 (2007) F685.

\bibitem{Widmer}
P.\ Widmer, Contact Interactions in Non-relativistic Quantum Mechanics, bachelor theses ( 2009).

\bibitem{Fas013}
S.\ Fassari, J.\ Phys.\ A:\ Math.\ Theor.\ 46 (2013) 385305.
\bibitem{Cal14}
M.\ Calcada, J.\ Lunardi, L.\ Manzoni, and W.\ Monteiro, arXiv:1404.0968.
\bibitem{Arnbak11}
H.\ Arnbak, P.\ Christiansen, and Y.\ Gaididei, Phil.\ Trans.\ R.\ Soc.\ A 369 (2011) 1228.
\bibitem{Fri73}
H.\ Fritzsch, M.\ Gell-Mann, and H.\ Leutwyler, Phys.\ Lett.\ 47B (1973) 365.

\bibitem{Gro73}
D.\ Gross and F.\ Wilczek, Phys.\ Rev.\ Lett.\ 30 (1973) 1343.

\bibitem{Pol73}
H.\ D.\ Politzer, Phys.\ Rev.\ Lett.\ 30 (1973) 1346.

\end{thebibliography}
\end{document}